\documentclass{article}

\usepackage{arxiv}
\usepackage{hyperref}
\usepackage[utf8]{inputenc} % allow utf-8 input
\usepackage[T1]{fontenc}    % use 8-bit T1 fonts
\usepackage{booktabs}       % professional-quality tables
\usepackage{amsfonts}       % blackboard math symbols
\usepackage{nicefrac}       % compact symbols for 1/2, etc.
\usepackage{microtype}      % microtypography
\usepackage{lipsum}         % Can be removed after putting your text content
\usepackage{graphicx}
\usepackage{natbib}
\usepackage{doi}
\usepackage{float} 
\usepackage{amssymb}
\usepackage{mathtools} 
\usepackage{multirow}
\usepackage{siunitx}
\usepackage{amsfonts} 
\usepackage{latexsym}
\usepackage{psfrag} 
\usepackage{url}
\usepackage{amsthm}
\usepackage[english]{babel}
\usepackage[shortlabels]{enumitem}
\usepackage{caption} \usepackage{subcaption}

\numberwithin{equation}{section}
\newcommand{\tr}{\mbox{tr}}

\theoremstyle{plain}% Theorem-like structures%
\newtheorem{lemma}{Lemma}[section]

\newtheorem{property}{Property}[section]

\theoremstyle{definition}

\theoremstyle{remark}
\newtheorem{remark}{Remark}[section]

\usepackage{cleveref}       % smart cross-referencing

\title{Measuring the severity of multi-collinearity in high dimensions}

\author{ \href{https://orcid.org/0000-0003-4212-2607}{\includegraphics[scale=0.06]{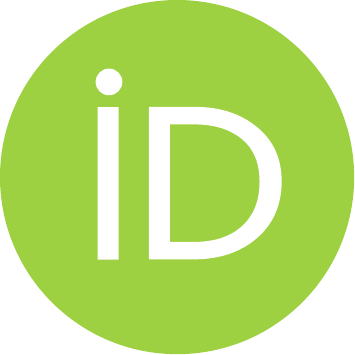}\hspace{1mm}Wei Q.~Deng} \\
Department of Psychiatry and Behavioural
Neurosciences\\
McMaster University\\
Peter Boris Centre for Addictions
Research\\
St. Joseph's Healthcare Hamilton\\
Hamilton, Canada\\
	\texttt{dengwq@mcmaster.ca} \\
	\And
	\hspace{1mm}Radu V.~Craiu \\
	Department of Statistical Sciences\\
	University of Toronto\\
	 Toronto, Canada \\
	\texttt{radu.craiu@utoronto.ca} \\
	\And
	\href{https://orcid.org/0000-0002-5640-937X}{\includegraphics[scale=0.06]{orcid.pdf}\hspace{1mm}Lei Sun} \\
	Department of Statistical Sciences\\
	Dalla Lana School of Public Health \\
	University of Toronto\\
	Toronto, Canada \\
	\texttt{sun@utstat.toronto.edu} \\
}

\renewcommand{\shorttitle}{New measures of multi-collinearity severity}

\hypersetup{
pdftitle={},
pdfauthor={Wei Q.~Deng, Radu V.~Craiu, Lei Sun},
pdfkeywords={First keyword, Second keyword, More},
}

\begin{document}
\maketitle

\begin{abstract}
Multi-collinearity is a
wide-spread phenomenon in modern statistical
applications and when ignored, can negatively
impact model selection and statistical inference.
Classic tools and measures that were developed for
``$n>p$'' data are not applicable nor interpretable
in the high-dimensional regime. Here we propose 1)
new individualized measures that can be used to
visualize patterns of multi-collinearity, and
subsequently 2) global measures to assess the
overall burden of multi-collinearity without
limiting the observed data dimensions. We applied
these measures to genomic applications to
investigate patterns of multi-collinearity in
genetic variations across individuals with diverse
ancestral backgrounds. The measures were able to
visually distinguish genomic regions of excessive
multi-collinearity and contrast the level of
multi-collinearity between different continental
populations.
\end{abstract}

% keywords can be removed

\section{Introduction}
In the current age of information, statisticians
often benefit from the ubiquitous capacity to
measure multiple features or covariates for each
unit in a sample. For example, large-scale genomic analyses typically measure tens of millions genetic features for tens of thousands of samples and proteomic assay can characterize thousands of circulating proteins for hundreds of individuals~\citep{suhre2021genetics}. A somewhat reversal of this fortune occurs when the number of units $n$ does not keep pace with the number of features $p$,
thus leading to the high-dimensional data matrices
$X \in R^{n\times p}$ for which $n<p$
\citep{donoho2000high}. For such data, the
multi-collinearity phenomenon, in which one feature
vector is highly correlated to linear combinations
of the remaining ones, is inevitable and often
produces damaging effects on model selection and
statistical inference~\citep{fan2008sure}.
However, assessing the severity of
multi-collinearity in high dimensions is not
straightforward as such a measure must factor in
both the number of variables involved as well as
the degree of collinearity among variables. The
absence of a severity measure in high-dimensional
settings seems disconnected from the fact that
variable selection remains pivotal in balancing
model accuracy and interpretability
\citep{george2000variable,wasserman2009high}.
Moreover, identifying which covariates are
correlated or even redundant can be as important
as finding a subset with high explanatory power.

There are some potential candidates for measuring
multi-collinearity in high dimensions. The
\textit{Red} indicator~\citep{kovacs2005new} has
been proposed to quantify the average level of
correlation in the data. An almost identical
quantity is the root mean square correlation over
all $p(p-1)/2$ pairs variables introduced in
\cite{efron2010correlated}, also a key component
in the approximation of covariance. These are
single number measures that do not point to any
specific variables, but cast light on the
appropriate next-steps. For example, they can
guide the implementation of regularization or
penalization techniques in the context of
high-dimensional linear regression, such as least
absolute shrinkage and selection operator
(lasso;~\citealp{santosa1986linear,lassoTib}) or elastic net
regularization~\citep{zou2005regularization}.

We can perhaps learn from the more accessible
scenario of $n > p$, where diagnostic measures to
assess severity of multi-collinearity have been
reliably used, especially in the context of linear
regression
\citep{farrar1967multicollinearity,marquaridt1970generalized,
belsley2014conditioning}. These measures fall into
two categories, one relying on a collection of
numbers measuring the impact or burden of
multi-collinearity on each individual variable,
and the other category that uses a single number
to summarize the severity of multi-collinearity of
all variables or a subset of the variables.

Examples of the former include a class of measures
that incorporate various functions of the
estimated coefficient of determination $R_j^2$
from linear regression models. In essence, this
type of measure leverages information on how well
the $j$th variable is explained by linear
combinations of the others as an indicator of the
severity of collinearity. Among them, the most
commonly used is the variance inflation factor
(\textit{VIF};~\citealp{marquaridt1970generalized}), defined by \[ \textit{VIF}_j =
\frac{1}{1-R_j^2}, \] intuitively interpreted as
the inflating factor for the variance of the
estimated regression coefficient for the $j$the variable. \textit{VIF} not only captures the degree of multi-collinearity for each variable, but also illustrates a direct
impact on inference in linear regression
models~\citep{fox1984linear}. Departing from
examining $X$ alone for multi-collinearity, a
corrected \textit{VIF} (denoted
CVIF; ~\citealp{curto2011corrected}) was proposed to
differentiate variables based on whether the
redundant information is predictive of the
response variable or not. The corrected
\textit{CVIF} is preferred over \textit{VIF} when
the redundancy among variables is unrelated to the
response variable~\citep{curto2011corrected}. These
individual-valued measures offer a mechanism to
remove variables implicated in near or perfect
multi-collinearity according to a pre-defined
threshold (e.g.\ $\textit{VIF}_j$ or
$\textit{CVIF}_j > 10$), and thus ensure
coefficient estimates of the remaining variables
using ordinary least square (OLS) are numerically
stable.

The second class of measures uses a single number
to summarize multi-collinearity. The most notable
being the condition number, defined by the ratio
of the largest and smallest singular values of a
data matrix~\citep{rice1966theory,
geurts1982contribution}. It is directly related to
the matrix solution of a linear system and
describes the degree to which the matrix $X^{T}X$
is ill-conditioned. For a scaled data matrix with
unit variance in each column, a value between 15
and 30 is considered moderately problematic and
severe if above 100~\citep{belsley2005regression}.
Closely related is the condition
index~\citep{belsley2014conditioning}, which is
defined by the square root of ratio of the largest
eigenvalue and each of the remaining eigenvalue of
$X^{T}X$. The number of condition indices above a
threshold further indicates the number of near or
perfect multi-collinear relationships in the data.
Other global measures include those examining the
determinant of $X^{T}X$, such as the
Farrar-Glauber test
statistic~\citep{farrar1967multicollinearity} that
evaluates a function of the determinant of
$X^{T}X$.

In practice, application of the two classes of
measures need not be mutually exclusive. In fact,
it has been shown that \textit{VIF}s are bounded
above by the squared condition
number~\citep{berk1977tolerance,salmeron2018variance}, implying that there could be
additional information in condition number that is
not captured by \textit{VIF}s. Indeed, sometimes
problematic variables are restricted to a
particular subset while their individual
\textit{VIF}s might not all be strong enough to be
picked up at the recommended threshold. A
generalization of the \textit{VIF} has been
proposed by~\cite{fox1992generalized} to measure
an arbitrary subset of variables for evidence of
multi-collinearity, which can be used to identify
specific sources of imprecision.

Under $n<p$, measures such as \textit{VIF} cannot
be reliably calculated, while overall measures
that rely on the sample eigenvalues or singular
values could be misleading, as we demonstrate in
Section~\ref{subsec:detection}. Further, the
usual approach to visualize pairwise relationship
quickly becomes cumbersome as the number of
combinations increases exponentially. Finally,
though the \textit{Red} indicator or the root mean
square correlation can be useful as an overall
summary, they do not fully address the complexity
of multi-collinearity in high-dimensional
settings.

This paper contributes in two new ways to the
study of multi-collinearity in the case of high-dimensional data. First, it introduces new measures for the severity of multi-collinearity derived via the
singular value decomposition (SVD) of $X$. Second,
it uses these novel measures to  establish whether
the multi-collinearity is due to all or just a few of the variables. The remaining paper is organized as follows. Section~\ref{sec:methods} introduces the individual-valued measures, presents their empirical properties, and motivates an overall summary measure. Section~\ref{sec:simulations} illustrates the utility of these measures to visualize and characterize multi-collinearity, making them an attractive option for exploratory data analysis on high-dimensional data. Section~\ref{sec:applications} demonstrates their application to genotype data from the 1000 Genomes Project~\citep{10002015global} to learn about the different patterns of multi-collinearity in genetic variations arising from diverse ancestral backgrounds.

\section{A severity measure of multi-collinearity}\label{sec:methods}

Let $X \in \mathbb{R}^{n \times p}$ be the
observed data matrix with each column standardized
to have sample mean 0 and variance 1. We are
interested in the high-dimensional data setting
($n < p$) that is the signature of large-scale
data such as those arising from genomic
applications, but results also naturally
generalize to the data rich setting ($n>p$).
Denote the SVD of $X$ by $UDV^{T}$, where columns
of $U \in \mathbb{R}^{n \times (n-1)}$ are the
left singular vectors, $D$ is a diagonal matrix
with singular values $d_1 \ge d_2 \ge \dots \ge
d_{n-1} \ge 0$, and columns of $V \in
\mathbb{R}^{p \times (n-1)}$ are the right
singular vectors. The column standardization
results in the loss of one degree of freedom such
that $\sum_{i'=1}^{n-1} d_{i'}^2=(n-1)p$, which is
the sum of the main diagonal elements of $X^{T}X
\in \mathbb{R}^{p \times p}$. Notice that by
permitting $d_{i'} = 0$, the matrix $X$ is allowed
to be rank deficient, which would be the
consequence of perfect collinearity involving two
or more variables.

Define the {\it right severity} measure of
multi-collinearity by \[ SR_j =
V_{j.}D^4V_{j.}^{T} =\sum_{i'=1}^{n-1}
v_{ji'}^2d_{i'}^4, \] where $V_{j.}$ denotes the
$j$th row and $v_{ji'}$ the $(j,i')$th entry of
$V$.

Naturally, the duality of SVD allows the
definition a {\it left severity}: \[ SL_i =
U_{i.}D^4U_{i.}^{T} = \sum_{i'=1}^{n-1}
u_{ii'}^2d_{i'}^4, \] where $U_{i.}$ denotes the
$i$th row and and $u_{ii'}$ the $(i,i')$th entry
of $U$.

Notice that these two measures are equal when $X$
is symmetric (i.e.\ $X = X^{T}$). Both $SR$ and
$SL$ leverage the spectrum of singular values of
$X$, similar to other measures of
multi-collinearity, but also the singular vectors,
which are used to assign a value to each
variable/sample through the weighted $l_2$ norm of
the corresponding right/left singular vector. In
this construction, the singular values
comprehensively capture the variance spectrum, and
weighting by their respective singular vectors
creates individualized measures irrespective of
the data dimensions. Since the top singular values
bear the higher burden of capturing the variance
in $X^{T}X$ and contribute more weight to the
measures, $SL_i$ and $SR_j$ are termed the
univariate burden of variance adjustment (uBVA)
measure for left and right severity, respectively.

\subsection{Basic
properties}\label{subsec:properties}

Without invoking any distributional or data
dimensions assumptions, we first establish three
basic properties of $\{SR_j\}_{j=1,\dots,p}$ given
any observed $X \in \mathbb{R}^{n \times p}$ with
each column standardized to have sample mean 0 and
variance 1.

\begin{property}\label{p1} \begin{equation}
\sum_{j=1}^p SR_j = \sum_{i=1}^n SL_i =
\sum_{i'=1}^{n-1} d_{i'}^4 \end{equation}
\end{property}

\begin{remark} Though the sums of $SR_j$ and
$SL_i$ are the same, the collective pattern of
these values is influenced by the underlying
column and row dependence, respectively.
\end{remark}

\begin{property}\label{p2} \begin{equation} SR_j =
\sum_{i'=1}^{n-1} v_{ji'}^2d_{i'}^4 =
(n-1)^2\sum_{j'=1}^{p}r_{jj'}^2, \end{equation}
\label{eq:22} where $r_{jj'}$ denotes the sample
Pearson's correlation coefficient between the
$j$th and $j'$th columns. Note that since the data
had been column-standardized, we have $r_{jj}^2 =
1$ for all $j = 1, \dots, p$.

\end{property}

\begin{remark} The equivalent expression of
$SR_j$, shown in equation~\eqref{eq:22}, offers
some intuition to the construction of the measure.
The magnitude of $SR_j$ scales with the variance
of the $j$th column itself as well as any
redundancy due to its correlation with all other
columns. The larger $SR_j$ is, the more the $j$th
column is involved in multi-collinearity,
quantified by the number and severity of these collinear
relationships. \end{remark}

\begin{remark} Since $r_{jj'}^2 \in [0,1]$, the
maximum value of $SR_j$ is bounded by $(n-1)^2p$,
while the minimum possible value is bounded by
$(n-1)^2$. These bounds apply to any $X \in
\mathbb{R}^{n\times p}$, irrespective of $n > p$
or $n < p$. However, a tighter bound is
established in the next property when we restrict
data dimensions to be $n < p$. \end{remark}

\begin{property}\label{p3} When $n < p$,
\begin{equation*} SR_j \in
\left[\frac{(n-1)^2}{\sum_{i'=1}^{n-1} v_{ji'}^2},
d_1^2(n-1)\right]. \end{equation*} \end{property}

\begin{remark} When $n > p$, the lower bound
becomes $(n-1)^2$ assuming all $p$ columns are
mutually orthogonal. However, when $n < p$, $X$
has at most $\text{min}(n, p)-1$ orthogonal
columns; the restriction of dimension ($n<p$)
leads to a tighter lower bound than $(n-1)^2$
because the squared row norm of a column orthogonal
matrix is strictly less than 1 (i.e.\
$\sum_{i'=1}^{n-1} v_{ji'}^2 < 1$). In fact,
following from $\sum_{j=1}^p \sum_{i'=1}^{n-1}
v_{ji'}^2 = n-1$, the lower bound
$\frac{(n-1)^2}{\sum_{i'=1}^{n-1} v_{ji'}^2}$ is
expected to vary for each $j$, but the smallest
such lower bound is strictly smaller than
$(n-1)p$. In other words, the smallest value
$SR_j$ can take under a high-dimensional data setting is
greater than that under the one with $n>p$, the
result of spurious correlation as discussed
in~\cite{fan2012variance}. Clearly, the severity
increases with an increasing $p/n$ ratio.
Meanwhile, as $\sum_{i'=1}^{n-1} d_{i'}^2 =
(n-1)p$, the upper bound $d_1^2(n-1)$ is also
bounded above by the naive upper bound of $(n-1)^2p$,
but these two are equivalent when columns of
$X$ are identical and $\sum_{i'=1}^{n-1} d_{i'}^2
= d_1^2$. \end{remark}

\begin{remark} Under column standardization, the
bounds of $SL_i$ are not directly informative as
the singular values are scaled to have unit
column variance. Thus, we provide bounds for
$SL_i$ assuming row standardization and $n < p$,
which implies that $\sum_{i'=1}^{n-1} d_{i'}^2 =
n(p-1)$ and $\sum_{i'=1}^{n-1} u_{ii'}^2d_{i'}^2 =
p-1$. The upper bound is then: \[ SL_i =
\sum_{i'=1}^{n-1} u_{ii'}^2d_{i'}^4 \le
d_{1}^2\sum_{i'=1}^{n-1} u_{ii'}^2d_{i'}^2 \le
d_{1}^2(p-1), \] and the lower bound follows from
the Cauchy-Schwarz inequality: \[ SL_i =
\sum_{i'=1}^{n-1} u_{ii'}^2d_{i'}^4 \ge
\Big(\sum_{i'=1}^{n-1}
u_{ii'}^2\Big)\Big(\sum_{i'=1}^{n-1}
u_{ii'}^2d_{i'}^2\Big)^2 \ge (p-1)^2. \]
\end{remark}

Thus far, we have not invoked any distributional
assumptions. By assuming each row of $X$ follows a
multivariate normal distribution, the expected
value of $SR_j$ can be expressed in terms of the
true covariance matrix and data dimensions when $n
> p-1$.

\begin{lemma} Suppose \textbf{rows} of $X \in
\mathbb{R}^{n \times p}$ are independent and
identically distributed (i.i.d) normal random
vectors, i.e.\ for $i \in \{1, \dots, n\}$, $x_i
\sim \mathcal{N}(0, \Sigma)$, where $\Sigma$ is
positive definite with rank $p$ and $\Sigma_{j}$
is the $j$ column of $\Sigma$, then
\begin{equation*} \text{E}(SR_j) =
(n-1)\Sigma_{jj}\text{tr}(\Sigma) +
n(n-1)\Sigma_{j}^{T}\Sigma_{j}.\label{eqn:
qR_expected} \end{equation*}\label{prop_expected}
\end{lemma}

This result suggests that the expected value of
the proposed measure $SR_j$ has two components,
one that is driven by data dimensions ($n$ and
$p$) and the other by the non-zero off-diagonal
entries in the corresponding columns of $\Sigma$.

\begin{remark} The above result does not apply to
the $n < p$ setting as the scaled sample
covariance $X^{T}X$ no longer follows a Wishart
distribution due to the insufficient degrees of
freedom~\citep{wishart1928generalised}. In this
case, $X^{T}X$ is said to have a singular Wishart
distribution and explicit moments are not
available~\citep{srivastava2003singular}. An
alternative solution is to consider a low-rank
approximation of $X^{T}X$ with rank $r$ ($r < n$)
and compute the approximated expectation, at the
cost of slightly underestimating $\text{E}(SR_j)$.
\end{remark}

% It is still possible to derive the expectation
% by partitioning $X{T}X$ at the true rank $r < n$
% as the submatrix with dimensions $r\times r$ has
% a Wishart distribution $\mathcal{W}_r(\Sigma,
% n)$ and the off-diagonal submatrices. However,
% an arbitrary choice of $r$ could bias the
% $\text{E}(SR_j)$ as the diagonal submatrix of
% dimension $(p-r)\times(p-r)$ would not be
% explicitly determined the other submatrices.
%This is because the row norm of the right
%singular vectors is always less than 1 (i.e.\
%$\sum_{i'=1}^{n-1} v_{ji}^2 \le 1$) when $n < p$,
%and under the normality assumption,
%$\text{E}(\sum_{i'=1}^{n-1} v_{ji}^2) =
%\frac{n}{p}$. 

Though the main focus here was on the empirical
properties of these measures without
distributional assumptions, it is possible to
further characterize the statistical properties of
$SR_j$ or $SL_i$ according to behaviours of the
singular values and vectors using random matrix
theory such as in~\cite{bai2008methodologies}.
This will be the subject of future work. 
%Forexample, given a matrix with entries that are independent identically distributed random variables having zero mean and unit variance, the Marchenko-Pastur law~\citep{marchenko1967distribution} establishes the upper limit on the largest singular value $d_1$ at $(1+\sqrt{n/p})\sqrt{p}$ and the lower limit on the smallest singular value $d_n$ at $(1-\sqrt{n/p})\sqrt{p}$ with high probabilities.

Following property~\ref{p2}, it is natural to
define a scaled measure: \begin{equation} sR_j =
\frac{SR_j}{(n-1)^2} = \sum_{j'=1}^{p}r_{jj'}^2
\in [1, p], \label{bounds_sR} \end{equation} as it
has a more natural interpretation of being the sum
of squared pairwise Pearson's correlation
coefficients. From the row perspective, $sL_i$ can
also be defined similarly, provided the data had
been row standardized: \begin{equation} sL_i =
\frac{SL_i}{(p-1)^2} \in [1, n]. \end{equation}

As the results in Section~\ref{subsec:properties}
can be conveniently expressed by a rescaling, the
bounds on $sR_j$ become: \begin{equation} sR_j \in
\left[\frac{1}{\sum_{i'=1}^{n-1} v_{ji'}^2},
\frac{d_1^2}{n-1}\right], \label{p2sRj}
\end{equation} where $\frac{1}{\sum_{i'=1}^{n-1}
v_{ji'}^2} \le 1$, taking equality when $n > p$;
and $\frac{d_1^2}{n-1} \le p$, taking equality
when all columns are identical (i.e.\
$\sum_{i'=1}^{n-1} d_{i'}^2 = d_1^2 = (n-1)p$).
The upper and lower bounds are expected to be
numerically close when multi-collinearity is
driven by spurious correlation due to $n < p$
alone, but further apart as both the number and
strength of multi-collinear relationships
increase.

The result in Lemma~\ref{prop_expected} becomes:
\begin{equation} \text{E}(sR_j) = \frac{p}{n-1} +
\frac{n}{n-1}\Sigma_j^{T}\Sigma_j, \label{exp_sR}
\end{equation} which reveals the direct impact of
relative data dimensions, $p/(n-1)$, on the
severity of multi-collinearity.

\subsection{sRs: a unifying measure of
multi-collinearity}\label{subsec:sRs}

Since $\{sR_j\}$ is considered the individualized
measure of multi-collinearity, we propose a
summary measure \textit{sRs} as a weighted sum of
$sR_j$ with two components: 
\begin{align}
\nonumber sRs & = \frac{\sum_{j=1}^p sR_j -
p}{p(p-1)} \times \frac{w_1+w_2}{2} +
\frac{\sum_{j=1}^p sR_j - p}{p[d_1^2(n-1)^{-1}-1]}
\times \Big(1-\frac{w_1+w_2}{2}\Big) \in [0, 1],
\label{def:sRs}\\ 
\end{align} 
where $$w_1 =
\frac{\sum_{d_i^2 > p} d_i^2}{\sum d_i^2}$$
adjusts the weight of `bulk'' behaviour more
heavily when $n<p$ and $$w_2 = \frac{\sum_{d_i^2 >
(\sqrt{n}-\sqrt{p})^2} d_i^{-2}}{\sum d_i^{-2}},$$
so that $1-w_2$ adjusts the weight of ``local''
behaviour more pronouncedly when $n > p$.

We refer to the first component in~\eqref{def:sRs}
as bulk sRs (\textit{BsRs}): \begin{equation}
\textit{BsRs} = \frac{\sum_{j=1}^p sR_j -
p}{p(p-1)}, \end{equation} which captures the
overall burden of multi-collinearity, weighted by
the proportion of singular values exceeding their
averaged value (maximum of $n$ or $p$). The second component in~\eqref{def:sRs} is
designed specifically to account for the number of
``locally'' strong relationships and defined as
local sRs (\textit{LsRs}): \begin{equation}
\textit{LsRs} = \frac{\sum_{j=1}^p sR_j -
p}{p[d_1^2(n-1)^{-1}-1]}. \end{equation}

Notice that, $p$ is used as the upper bound for
$sR_j$ when the ``bulk'' behaviour dominates,
i.e.\ the majority of variance is in the leading
singular values, while the maximum given
by~\eqref{p2sRj} is used when the top singular
values do not dominate others. In other words, the
combined measure includes each $sR_j$ but weighs
the signal relatively.

The \textit{sRs} ``bulk'' component is
mathematically equivalent to the squared
\textit{Red}, defined as \begin{equation}
\text{Red} = \sqrt{\frac{\text{tr}[X^{T}XX^{T}X
-(n-1)^2I_p]}{p(p-1)(n-1)^2}}, \end{equation} and
both describe the ``average correlation'' of all
variables. But the addition of a ``local
component'' in  \textit{sRs} helps account for
strong ``local'' collinearity that involves only a
subset of the variables.

The lower bound of $sRs = 0$ is achieved when $n >
p$ and columns of $X$ are mutually orthogonal; the
upper bound of $sRs = 1$ is achieved when columns
of $X$ are identical. Our proposed \textit{sRs},
along with \textit{LsRs} and \textit{BsRs}, have
better interpretation as compared to the
\textit{Red} indicator, e.g.\ a value closer to 0
suggests no evidence of multi-collinearity. In
contrast, a value closer to 1 indicates severe
multi-collinearity due to 1) a subset of variables
(local), a scenario that \textit{Red} was unable
to capture; 2) a large number of variables (bulk);
3) both 1) and 2). The relative contribution of
\textit{LsRs} and \textit{BsRs} to \textit{sRs}
can be used to suggest which one of the scenarios
constitutes the main driver of observed
multi-collinearity.

\section{On the use of right severity measure for
data exploratory analysis}\label{sec:simulations}

This section focuses on the utility of $sR_j$ and
\textit{sRs} (\textit{LsRs} and \textit{BsRs})
through simulation studies. In
section~\ref{subsec:visual}, we applied $sR_j$ to
high-dimensional data simulated under different
covariance structures to confirm basic properties
of $sR_j$ and to explore its use for initial data
analysis. In Section~\ref{subsec:detection}, we
compared \textit{sRs} (\textit{LsRs} and
\textit{BsRs}) with existing measures to assess
multi-collinearity in data generated under various
multi-collinearity patterns assuming either a
high-dimensional ($n < p$) or data rich
scenario ($n >p$).

\subsection{Visualizing data covariance
structure}\label{subsec:visual}

As each $sR_j$ is a weighted sum of singular
values and that the spectrum of singular values is
driven by the covariance structure from which the
data were sampled, it is tempting to use
$\{sR_j\}_{j=1,\dots,p}$ to identify certain
``signatures'' in the sample covariance through a
visual inspection. Property~\ref{p2} (i.e.\ $sR_j
= \sum_{j'=1}^{p}r_{jj'}^2$), suggests that the
observed range of $\{sR_j\}_{j=1,\dots,p}$ is
directly related to the number and strength of
squared pairwise correlation coefficients. While
Property~\ref{p3} implies that the observed
extremes of $\{sR_j\}_{j=1,\dots,p}$ are specific
to the data beyond dimensions. In summary, the
observed patterns of $sR_j$ are reflective of the
singular values and can be visualized to give a
fuller picture of the covariance structure.

With simulated examples, we demonstrate the
usefulness of this measure to differentiate some
representative covariance structures. The data
dimensions were fixed at $n = 500$ and $p =
1,000$. Following the standard notation, we use
$J_p$ to denote an $p\times p$ matrix of ones and
$I_p$ a $p\times p$ identity matrix. Each row of
$X$ was generated according to $x_i \sim
\mathcal{N}(0, \Sigma)$, where
\begin{enumerate}[A.] \item $\Sigma = I_p$ denotes
a case of identity covariance, \item $\Sigma =
J_p\rho + (1-\rho) I_p$, a compound symmetric
structure with $\rho = 0.2$, \item $\Sigma_{i,j} =
\rho^{|j-i|}$, a first order autoregressive (AR1)
structure with $\rho = 0.8$. \item $\Sigma =
\text{diag}[(0.1\times J_{p/2} + 0.9 I_{p/2}),
(0.4 \times J_{p/4} + 0.6 I_{p/4}), (0.6\times
1_{p/4} + 0.4 I_{p/4})]$, a covariance with three
compound symmetric blocks, \item $\Sigma = LL^{T}
+ \zeta^2 I_p$, a spiked covariance with two
distinct eigenvalues; the low-rank representation
$L = V_{,1:k}O$ is given by the first $k$ columns
of the right singular vectors, $k = 10$, $O =
\frac{1}{\sqrt{n}}\text{diag}[\sqrt{10}, \dots,
\sqrt{10}]$, and $\zeta^2 = 0.4$, \item $\Sigma =
LL^{T} + \zeta^2 I_p$, a spiked covariance with
$k+1$ distinct eigenvalues; $k = 10$, $O =
[o_1,\dots, o_k]$, where $o_k^2 = 2 + \zeta^2$ and
$\zeta^2 = 0.4$. 
\end{enumerate}

\begin{figure}
\includegraphics[width=0.92\linewidth]{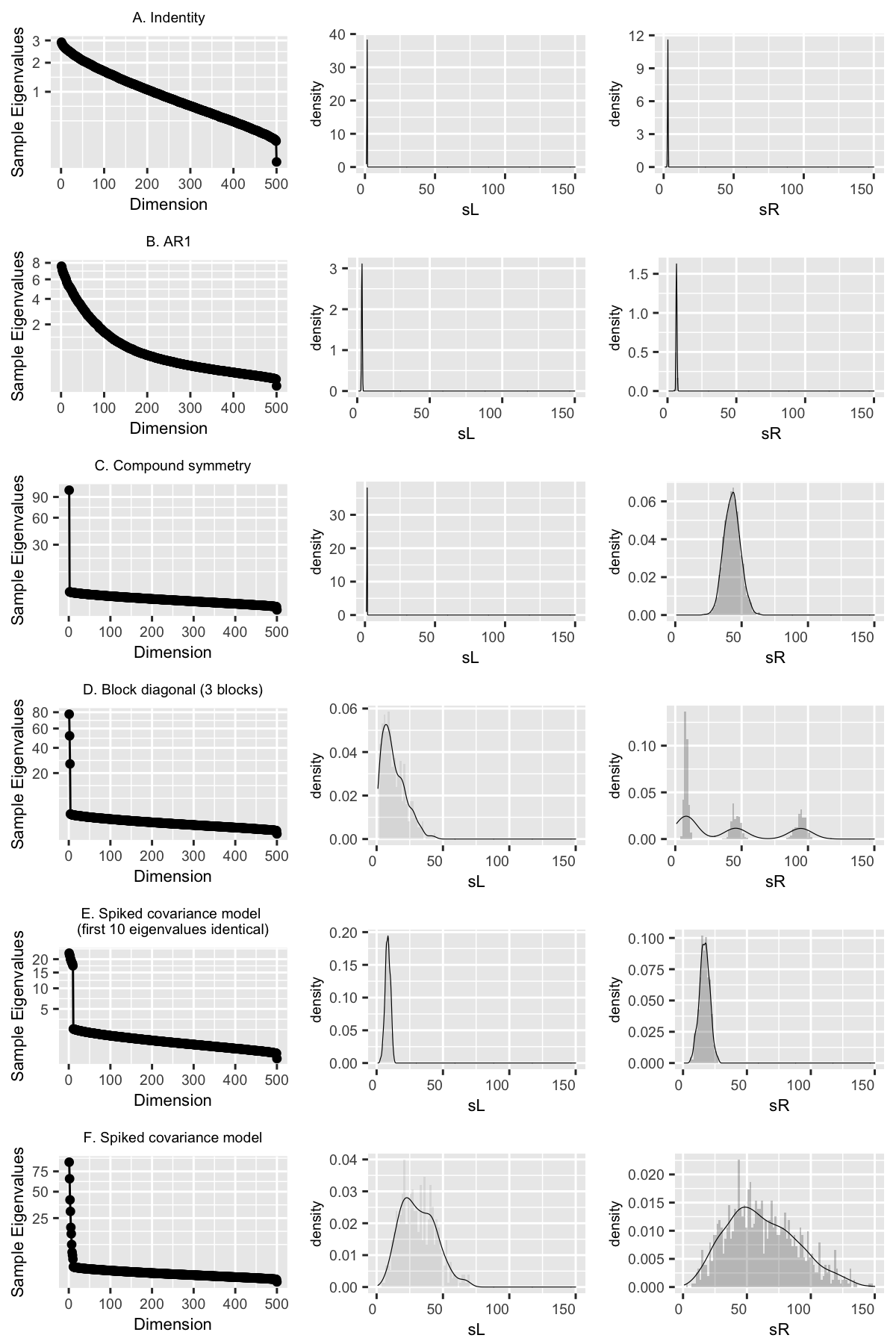} \caption{Empirical distributions of
sample eigenvalues, $sL_i$ and $sR_j$ under
different data covariance
structures.}\label{fig:Identify_covariance}
\end{figure}

For the spiked covariance models, when $O$ is not
given explicitly, we assumed $o_1^2, \dots,
o_{k-1}^2$ follow an exponential decay, which uniquely determined the values via the constraints imposed by $o_k^2 = 1 + \zeta^2$ and $\sum_{i=1}^k o_i^2 + p\zeta^2 = p$.

The sample eigenvalues (or normalized squared
singular values) shown in
Figure~\ref{fig:Identify_covariance} (A-F) have
distinct patterns under each structure: a
relatively smooth decay in the case of an identity
covariance (A) and AR1 structure (C); a sharp drop
is identified for the compound symmetry case (B),
the block-wise compound symmetric covariance (D),
and the spiked covariance with two identical true
eigenvalues (E); and finally a visible ``elbow''
for the spiked covariance with $k+1$ unique true
eigenvalues (F). However, these might not be
sufficient to differentiate the block-diagonal and
the low-rank spiked covariance cases. This is
where $\{sR_j\}$ could lend additional
information.

The identity case
(Figure~\ref{fig:Identify_covariance}-A) is
equivalent to each entry having a standard normal
distribution and Marchenko-Pastur
law~\citep{marchenko1967distribution} applies. We expected the observed $sR_j$ and $sL_i$ to fall between
$(\frac{p}{n}, (1+\sqrt{n/p})^2\frac{p}{n})$ and
$(1,\frac{d_{1}^{2}}{p})$, which translate to
$(2.00, 5.83)$ and $(1.00, 2.91)$, respectively.
These are consistent with the observed ranges of
$(2.72, 3.31)$ for $sR_j$ and $(1.15, 1.81)$ for
$sL_i$ (Figure~\ref{fig:Identify_covariance}-A).
The observed $sR_j$ had a tight symmetrical shape, with most values centred around its observed median
($3.005$), which was approximately the same as its observed mean ($3.007$).

The sample eigenvalues of an AR1 structure
(Figure~\ref{fig:Identify_covariance}-C) behaved
similarly to that of an identity covariance with
the additional variance for each principal
direction contributed by only nearby variables.
The empirical pattern should also be unimodal and
symmetric around its mean/median, but differ in the extremes from the identity case. In practice, for large $p$, the majority of $sR_j$ should have expected values close or equal to the maximum because $\rho^{|j-i|}$ diminishes quickly as $|j-i|$ increases. For fixed data dimensions, the observed range of $(3.65, 5.70)$ was attributed to the parameter value $\rho = 0.3$. As $\rho$
increases, the range will become wider with a
smaller minimum and a larger maximum.

When the true covariance matrix $\Sigma =
(1-\rho)I_p + 1_p\rho$ has a compound symmetric
structure (Figure~\ref{fig:Identify_covariance}-B), the
largest population singular value is
$\sqrt{n[1+(p-1)\rho]}$ and the remaining $n-1$
singular values are $\sqrt{(1-\rho)(p-1)}$. For
small $\rho$ values, $sR_j$ is influenced mostly
by the top singular values and their corresponding
singular vectors. Since all pairwise variables
have the same $2\times 2$ covariance, as expected,
the empirical distribution of $sR_j$ was roughly
symmetric with a unimodal shape that peaked around
the mean ($42.84$) and median ($42.81$).

When $\Sigma$ exhibits a block structure and each
block is compound symmetric: \[ \Sigma =
\begin{bmatrix} \Sigma_1 & 0 & 0 \\ 0 & \Sigma_2 &
0\\ 0 & 0 & \Sigma_3 \end{bmatrix}, \] the
empirical patterns of $\{sR_j\}_j$ should feature three
visible modes corresponding to each block
similarly described for a compound symmetric
structure (Figure~\ref{fig:Identify_covariance}-D). However, note that if any two blocks are identical, $sR_j$
would simply be duplicated for the identically
distributed variables in these two blocks. As a
result, two of the three modes would completely
overlap, forming a single mode. In general, the
number of modes corresponding to the number of
unique blocks while the within block pattern
depends on the structure of that block.

The last scenario focused on variables with
varying magnitudes of pairwise correlation such
that the true covariance followed a spiked structure
whereby $\Sigma = VO^2V^{T} + \zeta^2I_p$
(Figure~\ref{fig:Identify_covariance}-E,F). Though
challenging to estimate sample covariance
directly, the empirical patterns of $sR_j$ was
mostly be driven by the top singular values whose
true values are proportional to diagonal elements
of $V$. As a result, the empirical patterns spread
much wider (ranging from $8.13$ to $138.43$) and
no modes would be unambiguously identified unless
the top singular values were truly identical.
Indeed, when the true covariance has equal
eigenvalues, the observed $\{sR_j\}$ (ranging from
$5.76$ to $30.26$) can be made to resemble a
compound symmetry covariance by varying the two
unique eigenvalues. Nevertheless, it can be argued
that the compound symmetry covariance is actually
a special case of a spiked covariance model with
only one spike.

\subsection{Measuring the severity of multi-collinearity}\label{subsec:detection}

We have proposed $\{sR_j\}$ and \textit{sRs} as individual-valued and summary-level measures, respectively, to assess severity of multi-collinearity in high dimensions where existing measures fall short. Here we benchmark their performance against alternatives under high-dimensional settings ($n < p$) and data rich settings ($n > p$). Sample size of the simulated data was varied ($n = 100$ and $n = 500$), and the number of variables was fixed at $p = 1,000$ for the high-dimensional or $p = 50$ for the data rich scenarios.

In contrast to the previous simulation study of general covariance structures, we specified covariance matrix to represent no multi-collinearity via an identity matrix (orthogonal design), multi-collinearity through two near-collinear variables (collinear), a moderate level of multi-collinearity impacting all variables through a compound-symmetric covariance matrix (CS), a severe multi-collinearity impacting all variables through a spiked covariance model (spiked covariance), and the most severe case of nearly all variables are identical (almost perfect multi-collinear). To make a more interesting comparison, we also included a block-wise scenario where two variables are near-collinear but the remaining variables follow a compound-symmetric covariance structure. Similar to the simulations in Section~\ref{subsec:visual}, each row of $X$ was generated according to ${x}_i \sim \mathcal{N}({0}, \Sigma)$, where

\begin{enumerate}
\item $\Sigma = {I}_p$,
\item (local) $\Sigma = \text{diag}[0.99\times{1}_2 + \sqrt{1-0.99^2}{I}_2 , I_{p-2}]$,
\item (bulk) $\Sigma = \rho{1}_p + (1-\rho){I}_p$ with $\rho = 0.3$ or $\rho = 0.99$,
\item (bulk and local) $\Sigma = \text{diag}[0.99\times{1}_2 + \sqrt{1-0.99^2}{I}_2, \rho{1}_{p-2} + (1-\rho){I}_{p-2}]$ with $\rho = 0.3$,
\item (local) $\Sigma = LL^{T} + \zeta^2{I}_p$ with $k = 10$, $o_k^2 = 1 + \zeta^2$, and $\zeta^2 = 0.4$.
\end{enumerate}

The alternative measures include \textit{VIF}, the condition number, and the \textit{Red} indicator. As \textit{VIF} can only be sensibly applied when $n > p$, it was only included for comparisons in the data rich scenarios. The condition number is defined by the ratio of the largest and smallest singular values of a data matrix and describes the degree to which the matrix $X^{T}X$ is ill-conditioned. For the high-dimensional case, it was taken to be $\frac{d_1}{d_{n-1}}$ due to the column standardization; while for the data rich case, it was calculated as $\frac{d_1}{d_p}$. By design, the condition number captures the degree of multi-collinearity rather than the number of collinear relationships. In other words, it evaluates the worst case scenario and as a result, one perfect collinear relationship is all it takes to reach infinity, i.e.\ when $d_p$ or $d_{n-1}$ is exactly zero.

\begin{figure}[H]
\centering
\begin{subfigure}{1\textwidth}
  \includegraphics[width=1.02\linewidth]{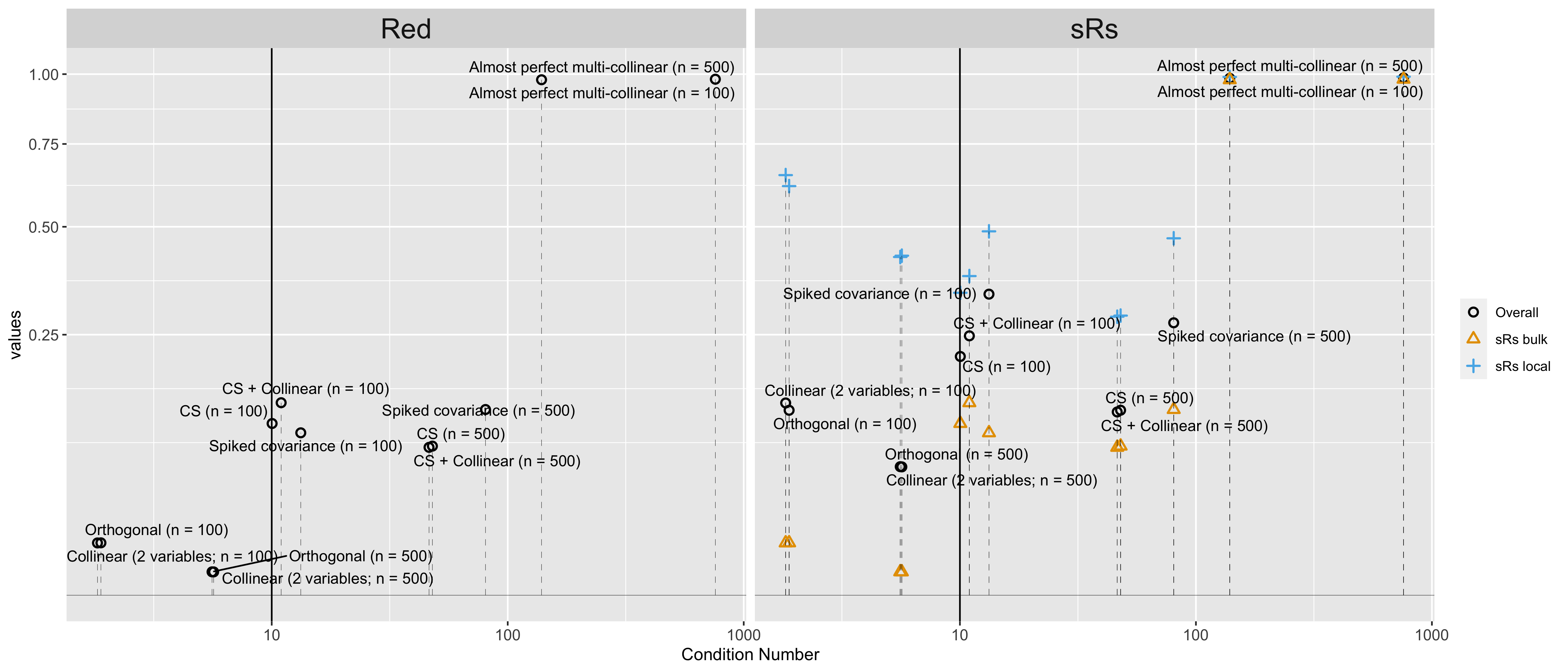}
  \caption{High-dimensional settings}\label{fig:highCompare}
\end{subfigure}

\begin{subfigure}{1\textwidth}
  \includegraphics[width=1.02\linewidth]{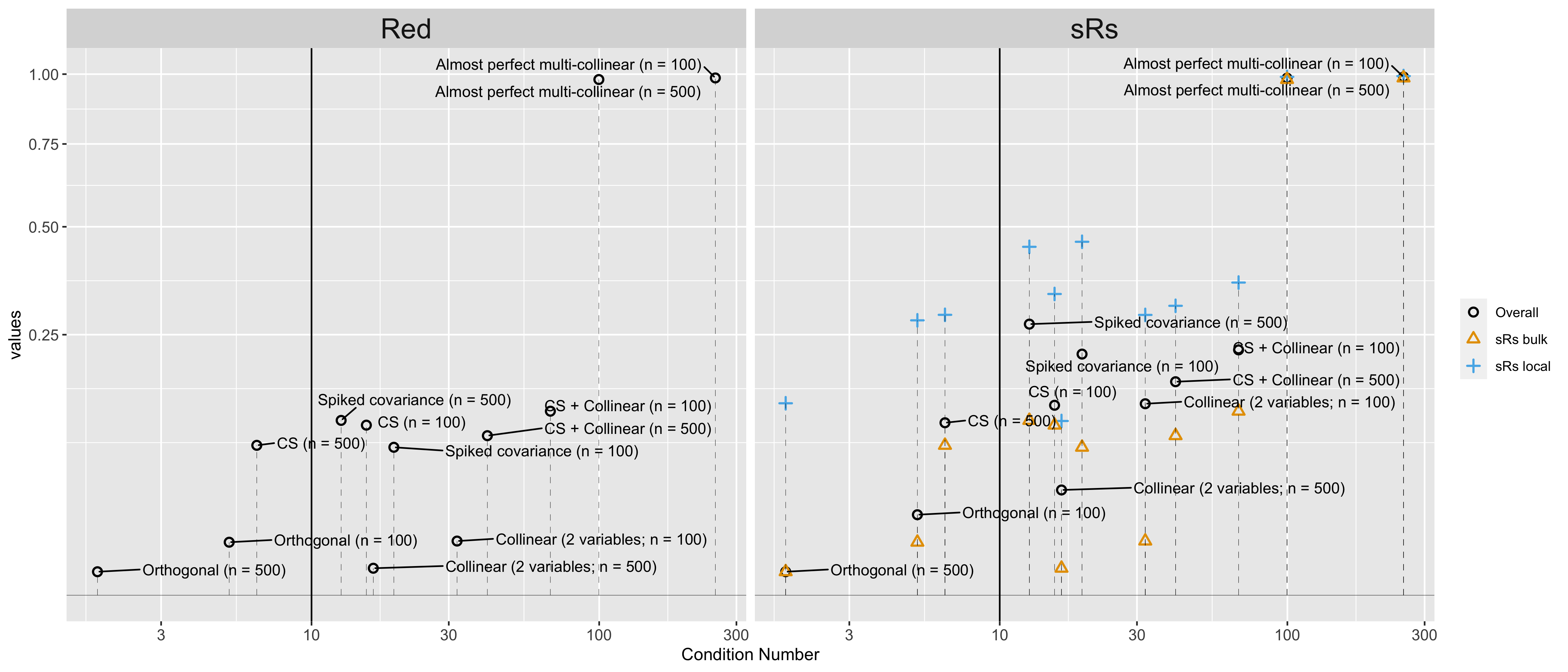}
  \caption{Data rich settings}\label{fig:lowCompare}
\end{subfigure}
\caption{Measuring overall severity of multi-collinearity using condition number, \textit{Red} indicator and \textit{sRs}. The vertical line marks the condition number cut-off at 10 to suggest presence of possible multi-collinearity.}\label{fig:compare}
\end{figure}

\paragraph{High-dimensional settings} We compared the \textit{Red} indicator and the proposed overall measure \textit{sRs} (Equation~\eqref{def:sRs}) to the condition number for assessing severity of multi-collinearity (Figure~\ref{fig:highCompare}). Unsurprisingly, the size of the condition number did not fully correspond to the severity of multi-collinearity under the impact of spurious correlations associated with high-dimensional data. In all scenarios, the \textit{Red} indicator was identical to \textit{BsRs}. But under $n <p$, when data matrices are necessarily under rank, and the \textit{LsRs} component received a boost through the trailing singular values that were very close to zero. Indeed, the main advantage \textit{sRs} had over the \textit{Red} indicator and condition number is its sensitivity to near-collinearity $\rho = 0.99$ due to the added \textit{LsRs} component. This contrast was expected since a global measure of ``averaged linear relationship'' might be less sensitive to local collinearity, for example, when two variables are near collinear. Given a fixed sample size, \textit{Red} ranked the compound symmetry structure to be less severely multi-collinear than the spiked covariance structure as opposed to the other way around for \textit{sRs}. This is because \textit{sRs} puts more weight on the bulk of correlations through their contributions to $d_1$, as well as locally strong correlation through their influences on $d_p$ or $d_{n-1}$. In the case of a low rank structure, the larger $sRs$ was due to the strong regional (between local and bulk) correlations, which contributed to the first a few leading singular values. In contrast, multi-collinearity under a compound symmetry structure has only one leading singular value, and thus the advantage of \textit{sRs} was less pronounced.

\paragraph{Data rich settings}

When $n > p$, \textit{sRs} showed better agreement with the commonly used condition number than \textit{Red} as reflected by points being closer to the line of reference, especially for the detection of two collinear variables (Figure~\ref{fig:lowCompare}). On the other hand, \textit{Red} was unable distinguish the scenarios of a compound symmetric covariance and that combined with two near collinear variables. Notice that a compound symmetric covariance with $\rho = 0.3$ is not considered to have a concerning level of multi-collinearity as these are $p$ very weak collinear relationships, all of the same size.

\begin{figure}[H]
\includegraphics[width=1.05\linewidth]{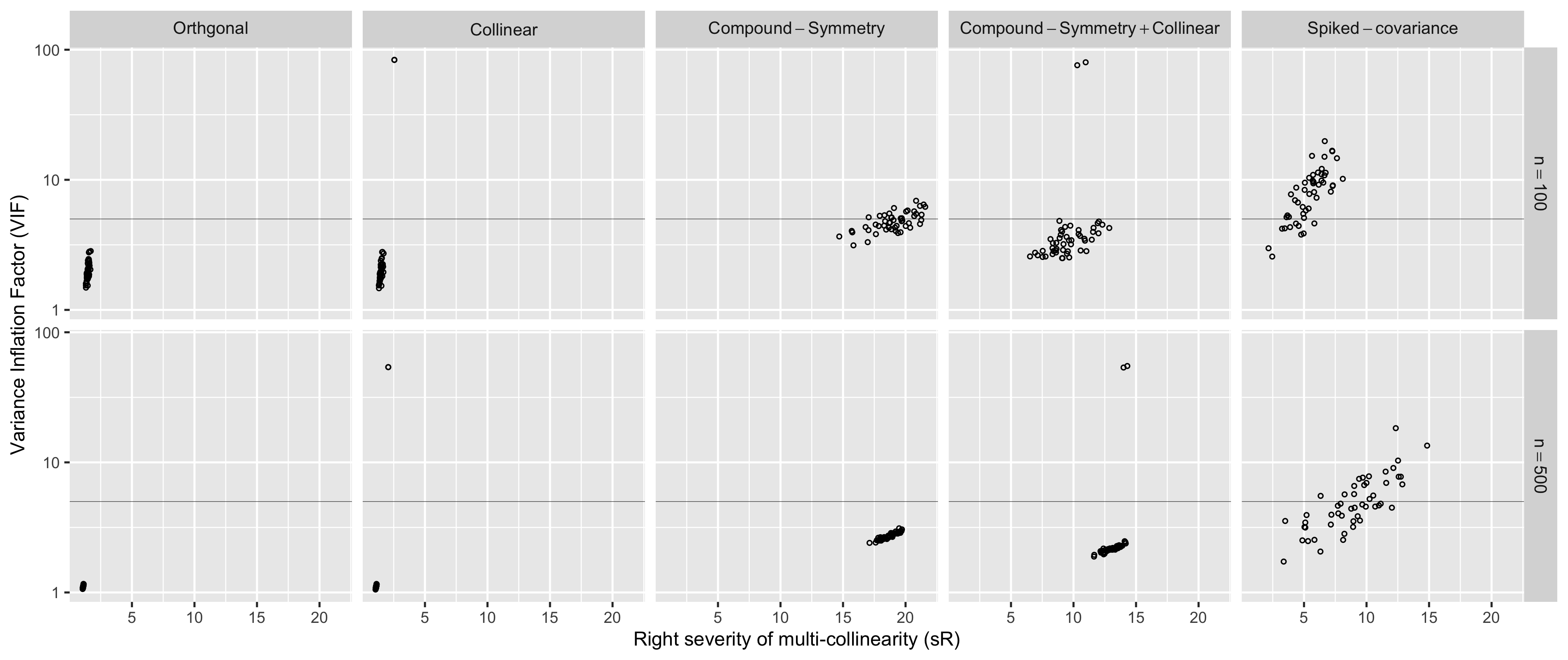}
\caption{A scatterplot of $\textit{VIF}_j$ and $sR_j$ under data rich settings. The horizontal line indicates a detection threshold of 5 for $\textit{VIF}_j$.}\label{fig:FigMultiCol}
\end{figure}

In terms of individualized measures, though $\textit{VIF}_j$ and $sR_j$ were designed to capture slightly different features of data, they did correlate to some extent, especially when sample size is large ($n = 500$; Figure~\ref{fig:FigMultiCol}). As discussed in Supplementary Section~\ref{sec:compareMeasure}, the numerical difference between the two measures is due to the joint correlation structure in the remaining $p-1$ variables. The results in fact suggested these two measures are complementary to each other. Given the same $\textit{VIF}_j$ value by varying the remaining $p-1$ variables, the pairwise correlation of $j$th variable with each of the $p-1$ variables can vary. For example, two variables having the same $\textit{VIF}_j$ value means they can be equally explained by the other $p-1$ variables. At the same time, the same two variables could have similar or very different $sR_j$ values, with a larger $sR_j$ suggesting the involvement of a larger number of individually weak relationships and a smaller $sR_j$ suggesting the involvement of a few, but stronger relationships.

\section{Application to the 1000 Genomes Project Data}\label{sec:applications}

The population genetics equivalent of multi-collinearity is linkage disequilibrium (LD), reflecting correlation between different genetic markers. For any pair of bi-allelic markers, the LD can be quantified by the squared Pearson's correlation coefficient. LD can be interpreted at the genome-wide scale to reflect population history, breeding system and the geographic subdivision within human populations~\citep{slatkin2008linkage}. At the same time, it can be viewed at a regional level indicating influences from selection, mutation and gene conversion~\citep{slatkin2008linkage}. Thus, as the number of genetic markers involved increased, the large numbers of pairwise Pearson's correlation coefficients make the studying of LD pattern over genomic regions of arbitrary size a challenging task.

The 1000 Genomes Project~\citep{10002015global} is a well-established reference for genetic variations and contains samples from several continental and sub-region populations. We applied the individual-valued uBVA and \textit{sRs}, along with \textit{LsRs} and \textit{BsRs}, measures to understand the severity of multi-collinearity within genetically homogeneous populations, as well as contrasting these measures across populations. The univariate $sR_j$ allows the comparison at each genetic marker, while the overall measures can be used to inform the overall burden of multi-collinearity.

\subsection{Data information and quality controls}

Standard quality controls on the genotype data are outlined in \cite{roslin2016quality} and the data are  publicly available (\url{http://www.tcag.ca/tools/1000genomes.html}). This set of data contains individuals from Africa (AFR; $n = 353$), East Asia (EAS; $n = 480$), Europe (EUR; $n = 522$), South Asia (SAS; $n = 100$), and the Latin America (AMR; $n=269$). The analyses were restricted to bi-alleleic markers on autosomes. For each continental population, we applied additional data filtering steps to exclude single nucleotide polymorphisms (SNPs) with minor allele frequency (MAF) less than 0.01, with any missingness, and Hardy-Weinberg Equilibrium $p$-value $< 1\mathrm{E}{-5}$. To harmonize the analysis in the combined sample, we retained only SNPs present in all continental populations, leaving 193,744 SNPs in the analysis, representing ~20\%-50\% of the SNPs originally available in each population. The genomic coordinates are based on the GRCh37/hg19 build. As a control step, we produced the first two genetic principal components using the subset of overlapping SNPs, and confirmed that they are sufficient to stratify samples at the continental and sub-population level (Figure~\ref{fig:PCA}).

\subsection{Chromosome specific patterns of multi-collinearity}

\begin{figure}
\includegraphics[width=1.1\linewidth]{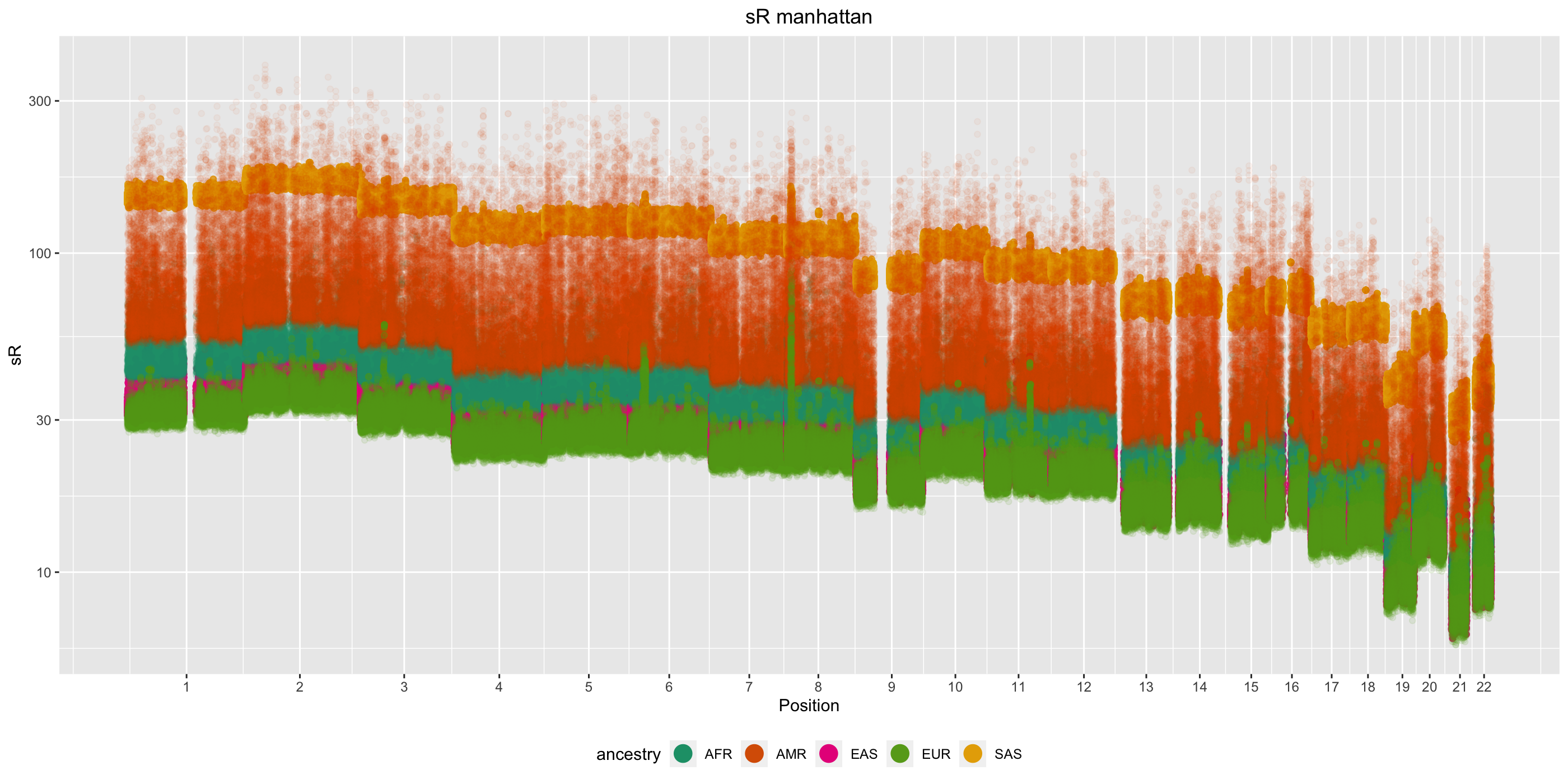}
\caption{A Manhattan type plot for $\{sR_j\}$ as a function of the genomic location within each chromosome.}
\label{fig:manhattan}
\end{figure}

The uBVA measures (i.e. $\{sR_j\}$) were calculated for each chromosome separately and presented in a similar manner to a Manhattan plot (Figure~\ref{fig:manhattan}), typical for genome-wide applications. We expected them to roughly follow the chromosome size (number of SNPs), but with varying peaks and valleys highlighting specific regions of high/low local multi-collinearity. Interestingly, there are a large number of visible peaks in the Europeans, only some of them are shared with other populations, in particular those well-known long range LD regions at chromosome 6, 8 and 11~\citep{price2008long} with slight differences in the exact location for different populations. For example, the region identified on chromosome 6 overlaps with the human leukocyte antigen (HLA) region. In Europeans, the peak region ranged from 25.9MB to 32.7MB, similar to the 26.4MB to 33.5MB in East Asians, but was much shorter in South Asians, only between 31.0MB to 32.3MB. 

In general, East Asian, European and South Asian have sporadic long-range LD regions, represented by the occasional peaks in Figure~\ref{fig:manhattan}, while populations in Latin America and Africa seemed to have more complex patterns of long and short range LD, corroborating the findings in \cite{park2019population}.

We observed two types of collective patterns of $\{sR_j\}$ across chromosomes and populations: those from populations in East Asia, Europe, and South Asian can be classified as being roughly symmetric (Figures~\ref{fig:EA_CHR},~\ref{fig:EU_CHR}, and~\ref{fig:SA_CHR}) and those from Africa and Latin America tended to have heavier tails for most chromosomes (Figures~\ref{fig:AMR_CHR},~\ref{fig:AFR_CHR}). The shift in the overall distribution should not be heavily influenced by outliers, such as the presence of a few long range LDs regions or strong LD blocks. Rather, considering the high-level of admixture in these populations, we hypothesized that these were probably the result of enriched genetic diversity manifested as a handful of large eigenvalues within each population (Figures~\ref{fig:AMR_summary},~\ref{fig:EA_summary},~\ref{fig:EU_summary},~\ref{fig:AFR_summary}, and ~\ref{fig:SA_summary}).

\subsection{Genome-wide summaries of multi-collinearity}

We then examined the overall level of multi-collinearity using genome-wide data (all 193,744 SNPs) and the results suggested the majority of multi-collinearity patterns were due to local relationships rather than global. The \textit{RED} indicator gave a slightly higher level of ``averaged correlation" in South Asia population and a lower level in Europeans and East Asians. On the other hand, \textit{sRs} offered better granularity in the type of forces driving the averaged correlation. Specifically, though both Europeans and East Asians had similar \textit{RED} values, their \textit{sRs} and \textit{LsRs} values collectively suggested a stronger local multi-collinearity in Europeans than in East Asians. 

% latex table generated in R 4.0.5 by xtable 1.8-4 package
% Wed Feb 16 07:53:50 2022
\begin{table}[ht]
\centering
\begin{tabular}{rrrrr}
  \hline
 & RED & sRs & LsRs & BsRs \\ 
  \hline
Combined & 0.01325 & 0.00180 & 0.00451 & 0.00018 \\ 
  SAS & 0.02365 & 0.01298 & 0.03674 & 0.00056 \\ 
  EAS & 0.01140 & 0.00396 & 0.01187 & 0.00013 \\ 
  AFR & 0.01323 & 0.00485 & 0.01465 & 0.00017 \\ 
  AMR & 0.01792 & 0.00363 & 0.00826 & 0.00032 \\ 
  EUR & 0.01092 & 0.00506 & 0.01608 & 0.00012 \\
   \hline
\end{tabular}
\caption{A genome-wide summary of multi-collinearity for each continental population and the combined samples.}
\end{table}

\section{Discussion}

Originally intended for detecting multi-collinearity under $n>p$, the \textit{Red} indicator is equally adaptable to rank the severity of multi-collinearity in high-dimensional settings ($n < p$). As a measure of ``averaged'' correlation in the data, \textit{Red} is sensitive to multi-collinearity that severely affects a large number of variables, but tends to ignore strong local relationships in the presence of moderate bulk relationships. Indeed, unlike \textit{sRs}, \textit{Red} does not have a mechanism to distinguish local and bulk relationships. On the other hand, the condition number is perhaps more specific to detect ill-posed problems as it is directly related to the numerical accuracy of the inverse of $X^{T}X$. Though these measures are sometimes useful as indicators of the overall multi-collinearity, they fall short in generating variable-specific information. As compared to $sR_j$, the individual-valued $\textit{VIF}_j$ can also identify specific variables involved in ill-conditioned problems and is capable of harvesting both local and global linear relationships, but can only be applied when $n < p$. In conclusion, the proposed \textit{sRs}, \textit{BsRs} and \textit{LsRs}, combined with $\{sR_j\}$ are recommended as measures of multi-collinearity in high-dimensional settings.

We want to highlight some potential improvements that are of interest for future research work. Firstly, $sR_j$ are empirical measures and a natural next step is to leverage theoretical results from random matrix theory to further derive their statistical properties. Secondly, it would be of interest to construct two-sample or multiple-sample statistical tests for quantities such as \textit{sRs}, \textit{BsRs} and \textit{LsRs}, thus enabling a formal statistical comparison of the severity and sources of multi-collinearity. Finally, though the application to autosomal markers yielded insightful results, the same might not translate to the X-chromosome due to differences in the number of chromosomal copies between sexes. One of the complications is that since females carry two copies and males carry only one copy of the X-chromosome, the multi-collinearity measures derived from the observed data are expected to vary with respect to the sex ratios. As a result, though the measures are still valid in the sense that they can be computed and reflect the observed severity of multi-collinearity, they cannot be reliably used to compare LD patterns between samples. 

It is worth noting that similar measures to $sR_j$ have been proposed in genetic applications: \textit{LDadj} was used in the construction of polygenic risk scores (PRS) for prediction~\citep{pare2016method,pare2017machine}, and \textit{LDscore} was used to demonstrate the polygenicity of a trait, such as in LD-score regression~\citep{bulik2015ld}. Both are in fact truncated versions of the $sR_j$ and are denoted by the sum of squared Pearson's correlation coefficients: 
\begin{equation}
\textit{LDadj}_{(j)} = \sum_{j'=j-t}^{j+t} r^2_{j'j},
\end{equation}
and 
\begin{equation}
\textit{LDscore}_{(j)} = \sum_{j'=j-t}^{j+t} r^2_{j'j} - \frac{1-r^2_{j'j}}{n-2},
\end{equation}
where \textit{LDscore} has an additional term such that each squared Pearson's correlation remains unbiased (under $n > t$). The value of $t$ is defined by assuming the neighbouring $t$ genetic variants up- and down-stream are sufficient to capture the local LD (or covariance) structure, thus the choice is subjective. A window of radius 1 centiMorgan around the index variant was recommended~\citep{bulik2015ld}. Contrary to uBVA, \textit{LDadj} or \textit{LDscore} do not intend to capture the systematic effect, but only the local effect of multi-collinearity. Consequently, these truncated measures are restricted to analyses within a homogeneous population and are not tailored for comparisons across samples of distinct populations; further, since the truncation occurs through a rolling window, the measures are not directly comparable from variable to variable.

\section{Concluding remarks}

Our multi-collinearity measures $\{sR_j\}$ offer an alternative univariate perspective to visualize multi-collinearity patterns. They also enable the construction of a high-level summary measure \textit{sRs} that sheds light on the sources of multi-collinearity through the relative contribution from \textit{LsRs} and \textit{BsRs}, which can inform the choice of an appropriate data learning strategy. The fact that these can be applied regardless of data dimensions is an attractive feature in high-dimensional data applications. Besides providing a visual inspection and numerical summary of multi-collinearity in high-dimensions, the proposed measures are amendable to various downstream analyses and potential applications, for example, as informative shrinkage weights to construct high-dimensional estimators. Finally, the simplicity in their construction also enables convenient data sharing for open science research.

\section*{Acknowledgement}
The authors would like to acknowledge Professors Dehan Kong and Stanislav Volgushev for a critical reading and their insightful comments on the early versions of the paper. The authors are grateful to Drs. Michael R. Chong and Nicholas Perrot for lengthy discussions and their constructive comments on the genetic applications.

\section*{Funding Information}
This research is funded by the Natural Sciences and Engineering Research Council of Canada (NSERC), the Department of Statistical Sciences at University of Toronto, and the Canadian Institutes of Health Research (CIHR).

\section*{Supplementary Materials}

\subsection*{Proofs}

\paragraph{Proof of property~\ref{p1}}

\proof:
Since $XX^{T}XX^{T}$ is a square matrix with non-negative eigenvalues $d_1^4, \dots, d_{n-1}^4$, the trace is simply the sum of eigenvalues:
\[
\tr(XX^{T}XX^{T}) = \sum_{i'=1}^{n-1} d_{i'}^4.
\]
It then follows from the cyclic property of trace that:
\[
\sum_{j=1}^p SR_j  = \tr(X^{T}XX^{T}X) =\tr(XX^{T}XX^{T}) = \tr(UD^4U^{T}) = \sum_{i=1}^n SL_i.
\]

\paragraph{Proof of property~\ref{p2}}

\proof:
\[
\text{diag}[\text{cor}(X)^2]_j  = \text{diag}\left[\frac{X^{T}XX^{T}X}{(n-1)^2}\right]_j = \frac{SR_j}{(n-1)^2},
\]
and
\[
\text{diag}[\text{cor}(X)^2]_j  = \sum_{j'=1}^{p}r_{jj'}^2.
\]

\paragraph{Proof of property~\ref{p2}}

\proof:
Following the Cauchy-Schwarz inequality, the lower bound is given by:
\begin{align}
\nonumber SR_j & = \sum_{i'=1}^{n-1} v_{ji'}^2d_{i'}^4 = \sum_{i'=1}^{n-1} \frac{(v_{ji'}^2d_{i'}^2)^2}{v_{ji'}^2} \\
\nonumber & \ge \frac{(\sum_{i'=1}^{n-1} v_{ji'}^2d_{i'}^2)^2}{\sum_{i'=1}^{n-1} v_{ji'}^2} \\
\nonumber & = \frac{(n-1)^2}{\sum_{i'=1}^{n-1} v_{ji'}^2} \ge (n-1)^2,
\end{align}
while the upper bound is:
\begin{align}
\nonumber SR_j & = \sum_{i'=1}^{n-1} v_{ji'}^2d_{i'}^4  \\
\nonumber & \le d_{1}^2\sum_{i'=1}^{n-1} v_{ji'}^2d_{i'}^2  \\
\nonumber & = d_{1}^2(n-1).
\end{align}

\paragraph{Proof of Lemma~\ref{prop_expected}}

\begin{proof}
Since $SR_j$ is simply the $j$th diagonal element of $X^{T}XX^{T}X$, we approach this by calculating the expected value of $SR_j = e_j^{T}X^{T}XX^{T}Xe_j$, where $e_j = (0,\dots, 0, 1, 0,\dots,0) \in \mathbb{R}^{p}$ is a standard basis vector with value 1 at the $j$th place. Further, it follows that $(n-1)\hat{\Sigma} = X^{T}X \sim \mathcal{W}_p(\Sigma, n)$ has a Wishart distribution with parameters $\Sigma$ and degrees of freedom $n$.

Following Proposition S1 of \citep{dicker2014variance}, where explicit expressions for the expectation of moments of a Wishart random matrix were derived, it is easy to write down the expectation of $SR_j$ as:
\begin{align}
\nonumber \text{E}(e_j^{T}X^{T}XX^{T}Xe_j) &= p(n-1)\frac{\text{tr}(\Sigma_p)}{p}e_j^{T}\Sigma_pe_j + n(n-1) e_j^{T}\Sigma_p\Sigma_p e_j \\
\nonumber & = (n-1)(\Sigma_p)_{jj}\text{tr}(\Sigma_p) + n(n-1)(\Sigma_p)_{.j}^{T}(\Sigma_p)_{.j}.
\end{align}
The expectation can be further simplified if the true covariance $\Sigma$ has diagonal elements 1:
\begin{equation*}
\text{E}(SR_j) = (n-1)p + n(n-1)\Sigma_j^{T}\Sigma_j. 
\end{equation*}
\end{proof}

\subsection*{Relationship with existing measures of multi-collinearity}\label{sec:compareMeasure}

Since measures are often derived from sample eigenvalues, which are closely related to singular values, here we reveal the relationship between proposed and existing measures of multi-collinearity.

\paragraph{The \textit{Red} indicator}

Both the \textit{Red} indicator~\citep{kovacs2005new} and $sR_j$ can be used without dimension restrictions and it turns out the two are closely related:
\[
\text{Red} = \sqrt{\frac{\text{tr}[X^{T}XX^{T}X -
(n-1)^2I_p]}{p(p-1)(n-1)^2}} = \sqrt{\frac{\sum_{j=1}^p sR_j - p}{p(p-1)}},
\]
where \textit{Red}~$\in [0, 1]$ and $sR_j \in [1, p]$ following Equation~\eqref{bounds_sR}. It is regarded as a global measure of average correlation in the data over all pairwise variables or the proportion of redundant information, with values closer to 1 indicating a large number of near or perfect multi-collinearity relationships and values closer to 0 indicating little evidence of multi-collinearity. Given how \textit{Red} is defined, the authors did not provide a recommended threshold at which a concerning level of multi-collinearity is present. 

In relation to \textit{Red}, $sR_j$ can be seen as its individual-level counterpart and is potentially more useful for contrasting variables for their relative involvement in multi-collinearity. It should emphasized that individual $sR_j$ values alone cannot distinguish between the ``bulk weak'' and ``local strong'' scenarios as the same $sR_j$ value could be given by many weak relationships or a few strong relationships. In reality, the ambiguity also remains for \textit{Red} indicator values that are closer to the middle. For example, a \textit{Red} value of 0.4 can be achieved by either a large number of weak collinear relationships or a small number of perfect or near collinear relationships, with the latter having a bigger impact on matrix solutions to linear regression problems.

\paragraph{Variance Inflation Factor (\textit{VIF})}

Though the \textit{VIF} is restricted to the setting of $n > p$, it is still of interest to compare $sR_j$ and $\textit{VIF}_j$ on an equal footing as individual-valued measures. Note that when $n>p$, components of the SVD of $X$ have different dimensions: ${U} \in \mathbb{R}^{n\times p}$, ${V} \in \mathbb{R}^{p\times p}$, and ${D} = \text{diag}[d_1, \dots, d_p] \in \mathbb{R}^{p\times p}$. In this case, $d_p$ does not equal to 0 following the column-wise mean and variance standardization. It should be noted that $\textit{VIF}_j$ is only suitable when the data matrix is full rank, while $sR_j$ can be calculated without such restriction.

The \textit{VIF} of the $j$th predictor can be expressed as:
\begin{align}
\nonumber \text{\textit{VIF}}_j & = \frac{1}{1-R_j^2} \\
\nonumber & = \frac{x_j^{T}x_j}{x_j^{T}x_j - x_j^{T}X_{-j}(X_{-j}^{T}X_{-j})^{-1}X_{-j}^{T}x_{j}} \\
& = \frac{1}{1 - (n-1)b_{-j}^{T}(X_{-j}^{T}X_{-j})^{-1}b_{-j}},
\end{align}
where $x_{j}$ denotes the $j$th column of $X$, $X_{-j}$ the data matrix with $j$th column removed, and $b_{-j} = \frac{1}{n-1}X_{-j}^{T}x_{j}$ the vectorized univariate regression coefficients estimated between the $j$th variable and each of the other $p-1$ variables. Using the same notation, the multivariate regression coefficients estimated using the $j$th variable as the response and the other $p-1$ variables as the predictors can be expressed as $(n-1)(X_{-j}^{T}X_{-j})^{-1}b_{-j}$.

The proposed measure $sR_j$ can be similarly expressed:
\begin{align}
\nonumber sR_j & = \sum_{j'=1}^{p}r_{jj'}^2 \\
\nonumber & = \Big[1 + \sum_{j'\ne j} (\frac{1}{n-1}x_{j'}^{T}x_{j})^2\Big]\\
& = (1 + b_{-j}^{T}b_{-j}).
\end{align}

Both $\textit{VIF}_j$ and $sR_j$ are driven by $b_{-j}$, with the main difference being how $b_{-j}^{T}b_{-j}$ is weighted. Note that as $(X_{-j}^{T}X_{-j})^{-1}$ is capable of simultaneously modelling relationship among the other $p-1$ variables, $\textit{VIF}_j$ is expected to be more sensitive than $sR_j$ at recognizing multi-collinearity that involves a large number of variables as each element of $b_{-j}^{T}b_{-j}$ merely describes the strength of a bivariate relationship. From the alternative expression of $sR_j$ according to the definition via the singular values, we obtain
\begin{equation}
sR_j = (n-1)^{-2}\sum_{i'=1}^p v_{ji'}^2d_{i'}^4  = d_1^2 (n-1)^{-2}\left[\sum_{i'=1}^p (v_{ji'}^2d_{i'}^2)\frac{d_{i'}^2}{d_1^2}\right].
\end{equation}
Though each $d_{i'}$ is weighted towards $sR_j$, the collective behaviour of $sR_j$ will be influenced by a large $d_1$ and therefore captures information in the condition indices $\{\frac{d_1}{d_{i'}}\}_{i'=1, \dots,p}$.

An important aspect is the detection of variables involved in multi-collinearity, which often requires a hard detection threshold. For the individual $sR_j$, suppose the data matrix is column standardized, one possible threshold for $sR_j$ could be due to property~\ref{p2} combined with an approximated distribution for the sample Pearson's correlation coefficient given in~\cite{stuart1994kendall}:
\[
r = \frac{t}{\sqrt{n-2+t^2}},
\]
where $t$ is a random variable following Student's \textit{t}-distribution with degrees of freedom $n-2$. This results holds approximately for large enough $n$ and the pairs of variables are assumed to be uncorrelated. It can be shown that $r^2$ then follows a beta distribution with shape parameters $1/2$ and $(n-2)/2$, and $\text{E}(r^2) = (n-1)^{-1}$. Thus, a possible threshold for departure from orthogonal columns using $sR_j$ could be $\frac{p-1}{n-1} + 1$ by summing up the $p-1$ expected values of squared sample Pearson's correlation coefficients assuming the true pairwise correlation is zero throughout.

\subsection*{Connection to the effective sample size and effective number of variables}\label{sub:df}

Typically, in a regression, correlated samples do not change mean estimation, but rather influence inference through increased variance. As a result, the same estimator under correlated samples should have a variance adjusted for the effective sample size. For correlated variables, an analogous concept is the effective number of variables, which serves as an upper bound for the effective degrees of freedom of a model (usually defined as the trace of the hat matrix connecting the response to its fitted values, e.g.\ $H = X(X^{T}X)^{-1}X^{T}$, for OLS regression). 

Here we focus on the effective number of variables and the effective sample size, without referencing a model fitting procedure, and show that the proposed measure of multi-collinearity can be used to inform the maximum possible values for both. As a result of the dual (i.e.\ column and row) perspectives on a data matrix $X$, the same technique can be applied to either $X$ or $X^{T}$ provided that the respective columns or rows are standardized to have mean 0 and variance 1. 

Given a row standardized $X$ and $n > p$, the effective sample size as determined by the left severity measure is at most:
\[
\sum_{i=1}^n \frac{1}{sL_i} \le \sum_{i=1}^n \sum_{i'=1}^{p} u_{ii'}^2 = p.
\]

Analogously, given a column standardized $X$ and $n < p$, $\sum_{j=1}^p \frac{1}{sR_j}$ can be considered the effective number of variables. Further, it can be shown that $\sum_{j=1}^p \frac{1}{sR_j}$ is at most $n-1$ following property~\ref{p3}. To see this, for each $j$:
\[
\frac{1}{sR_j} \le \sum_{i'=1}^{n-1} v_{ji'}^2,
\]
which suggests that:
\[
\sum_{j=1}^p \frac{1}{sR_j} \le \sum_{j=1}^p \sum_{i'=1}^{n-1} v_{ji'}^2 = n-1.
\]

Since $\frac{1}{sR_j}$ is a constant between $1/p$ and 1, it can be viewed as the amount of non-redundant information in a variable prior to model selection. Consider the extreme case when all variables were truly uncorrelated, but under the impact of spurious correlation in high dimensions, the maximum degrees of freedom becomes $\text{min}(n, p)-1=n-1$, meaning every variable is equally important prior to model selection. The opposite scenario is when all variables completely correlate with each other, the maximum degrees of freedom reduce to 1. Though each variable is equally important, their relative importance would be scaled by $\frac{1}{p}$, meaning the model can include any one of the variables.

However, these two concepts are really two sides of the same coin arising from the execution of a row or column standardization. For convenience, briefly consider a doubly centred ($\sum_{i=1}^n x_{ij} = \sum_{j=1}^p x_{ij} = 0$) and doubly standardized ($\sum_{i=1}^n x_{ij}^2 = n-1$ and $\sum_{j=1}^p x_{ij}^2 = p-1$) data matrix $X$. 

Suppose $n > p$, assume a multivariate normal model for each row of $X$ with independent samples:
\[
x_i \stackrel{\text{iid}}{\sim} \mathcal{N}(0, \Sigma), \quad i = 1, 2, \ldots, n,
\]
where $\Sigma \in \mathbb{R}^{p\times p}$. With the means removed, it follows that $\hat{\Sigma} = \frac{1}{n}X^{T}X$ has a scaled Wishart distribution with mean and variance 
\[
\text{E}(\hat{\Sigma}) = \Sigma \quad \text{and} \quad \text{Var}(\hat{\Sigma}) =  \frac{1}{n}\Sigma^{(2)},
\]
where $\Sigma^{(2)}_{jk, lh} = \Sigma^{(2)}_{jl}\Sigma^{(2)}_{kh} + \Sigma^{(2)}_{jh}\Sigma^{(2)}_{kl}$ and $\Sigma^{(2)} \in \mathbb{R}^{p^2 \times p^2}$. 

Similarly, suppose $n < p$, a multivariate normal model can be assumed for each column of $X$ with independent variables:
\[
x_j \stackrel{\text{iid}}{\sim} \mathcal{N}(0, \Phi), \quad j = 1, 2, \ldots, p,
\]
where $\Phi \in \mathbb{R}^{n\times n}$. It follows that $\hat{\Phi} = \frac{1}{p}XX^{T}$ has a scaled Wishart distribution with mean and variance 
\[
\text{E}(\hat{\Phi}) = \Sigma \quad \text{and} \quad \text{Var}(\hat{\Phi}) =  \frac{1}{p}\Phi^{(2)},
\]
where $\Phi^{(2)}_{jk, lh} = \Phi^{(2)}_{jl}\Phi^{(2)}_{kh} + \Phi^{(2)}_{jh}\Phi^{(2)}_{kl}$ and $\Phi^{(2)} \in \mathbb{R}^{n^2 \times n^2}$. 

Following Theorem 8.4 of \cite{efron2012large}, when rows of $X$ are not independent and $n > p$, the effective sample size $n_\text{eff}$ is defined by
\begin{equation}
n_\text{eff} = \frac{n}{1+(n-1)\Big[\frac{n\sum_{i'=1}^{p}d_{i'}^4}{n(n-1)p^2} - \frac{1}{n-1}\Big]} = \frac{n^2p^2}{\sum_{i'=1}^{p}d_{i'}^4} = \frac{n^2}{\sum_{i=1}^{n} sL_i}.
\end{equation}

In comparison, when columns of $X$ are not independent and $n < p$, the effective number of variables ($\text{p}_\text{eff}$) is defined by
\begin{equation}
\text{p}_\text{eff} = \frac{p}{1+(p-1)\Big[\frac{p\sum_{i'=1}^{n-1}d_{i'}^4}{n^2p(p-1)} - \frac{1}{p-1}\Big]} = \frac{n^2p^2}{\sum_{i'=1}^{n-1}d_{i'}^4} = \frac{p^2}{\sum_{j=1}^{p} sR_j}.
\end{equation}

The inequality of arithmetic and geometric means implies
\[
\frac{p^2}{\sum_{j=1}^{p} sR_j} \le \sum_{j=1}^{p} \frac{1}{sR_j},
\]
and 
\[
\frac{n^2}{\sum_{i=1}^{n} sL_i} \le \sum_{i=1}^{n} \frac{1}{sL_i},
\]
which shows that $\sum_{j=1}^{p} \frac{1}{sR_j}$ and $\sum_{i=1}^{n} \frac{1}{sL_i}$ are indeed the maximum possible values for the effective number of variables and effective sample size, respectively.

\newcommand{\beginsupplement}{%
        \setcounter{table}{0}
        \renewcommand{\thetable}{S\arabic{table}}%
        \setcounter{figure}{0}
        \renewcommand{\thefigure}{S\arabic{figure}}%
     }
\beginsupplement

\subsection*{Supplementary Figures}

\begin{figure}[H]
\includegraphics[width=0.93\linewidth]{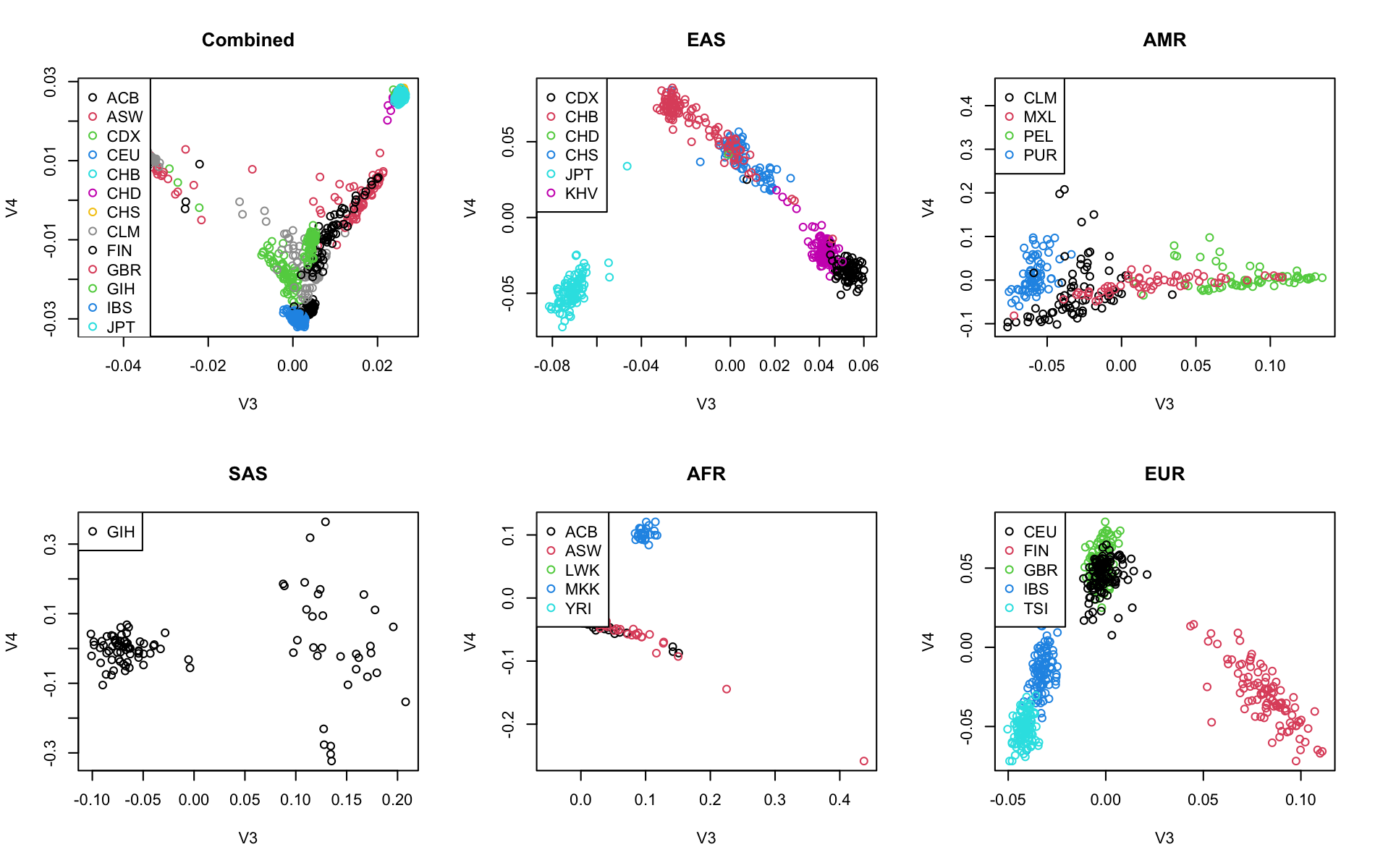}
\caption{Scatterplots of the first genetic principal components for the combined and each continental populations.}
\label{fig:PCA}
\end{figure}

\begin{figure}[H]
\includegraphics[width=0.93\linewidth]{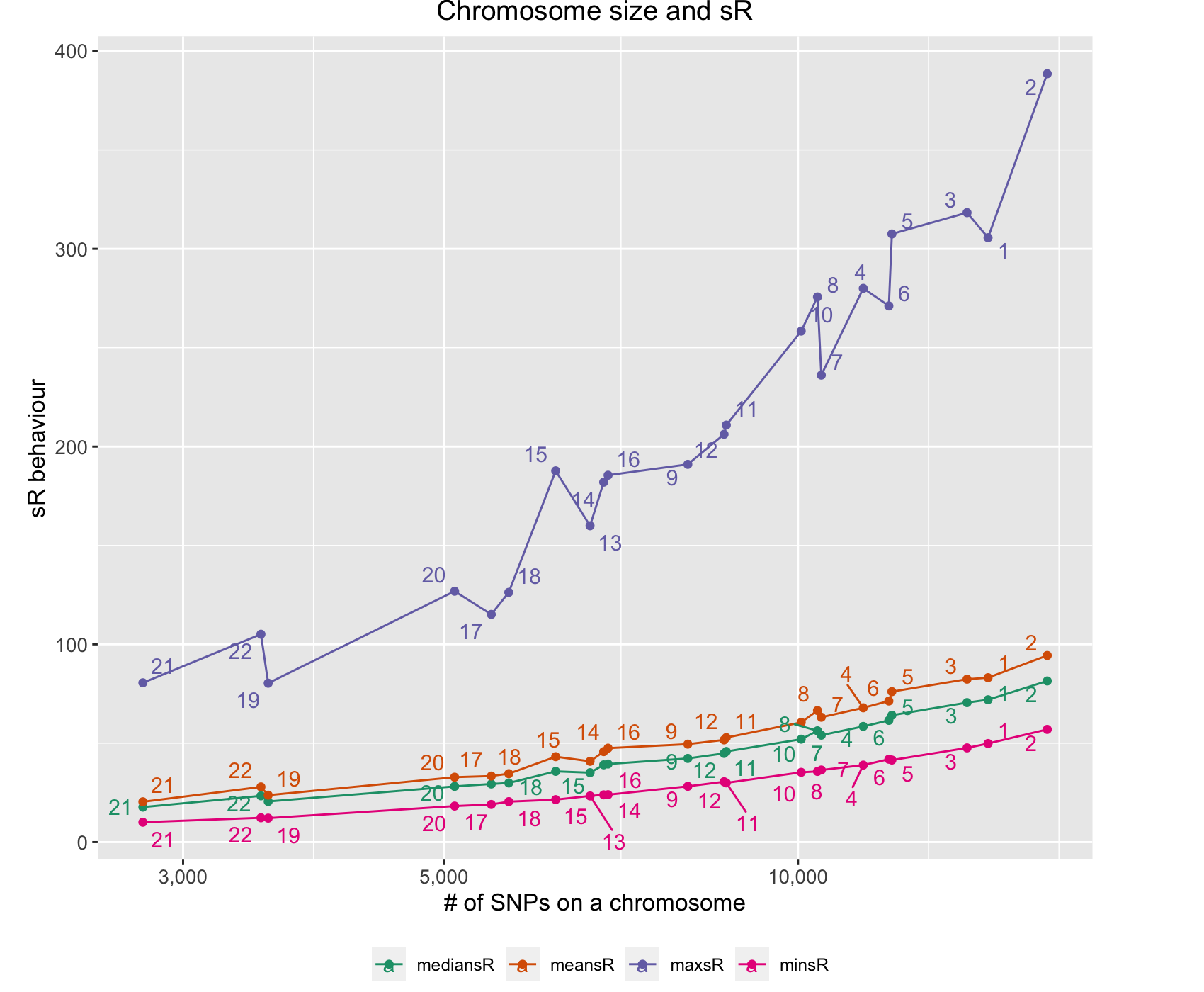}
\caption{A summary of multi-collinearity as a function of chromosome size for populations in America.}
\label{fig:AMR_CHR}
\end{figure}

\begin{figure}[H]
\includegraphics[width=0.93\linewidth]{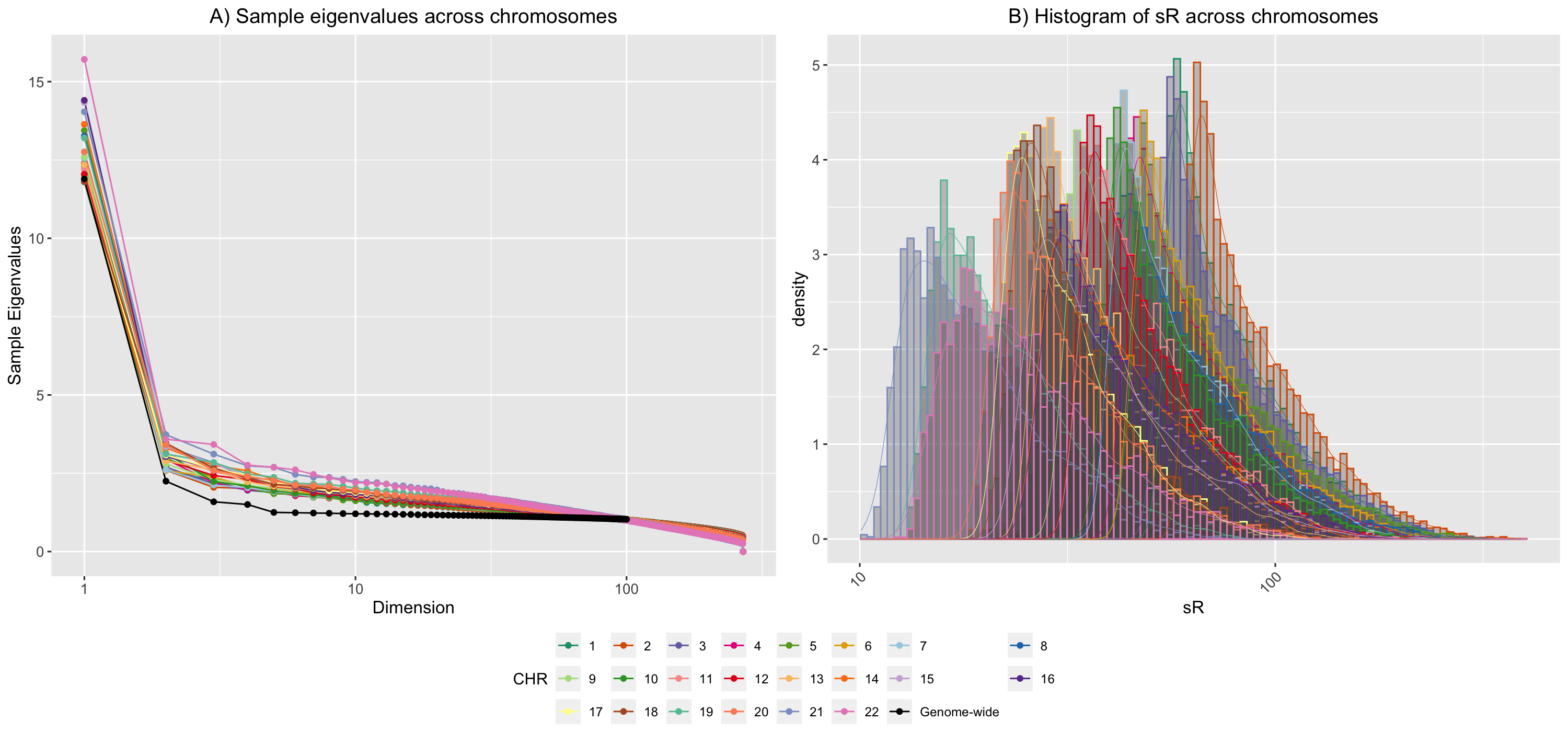}
\caption{Patterns of multi-collinearity measured by $\{sR_j\}_j$ in America.}
\label{fig:AMR_summary}
\end{figure}

\begin{figure}[H]
\includegraphics[width=0.93\linewidth]{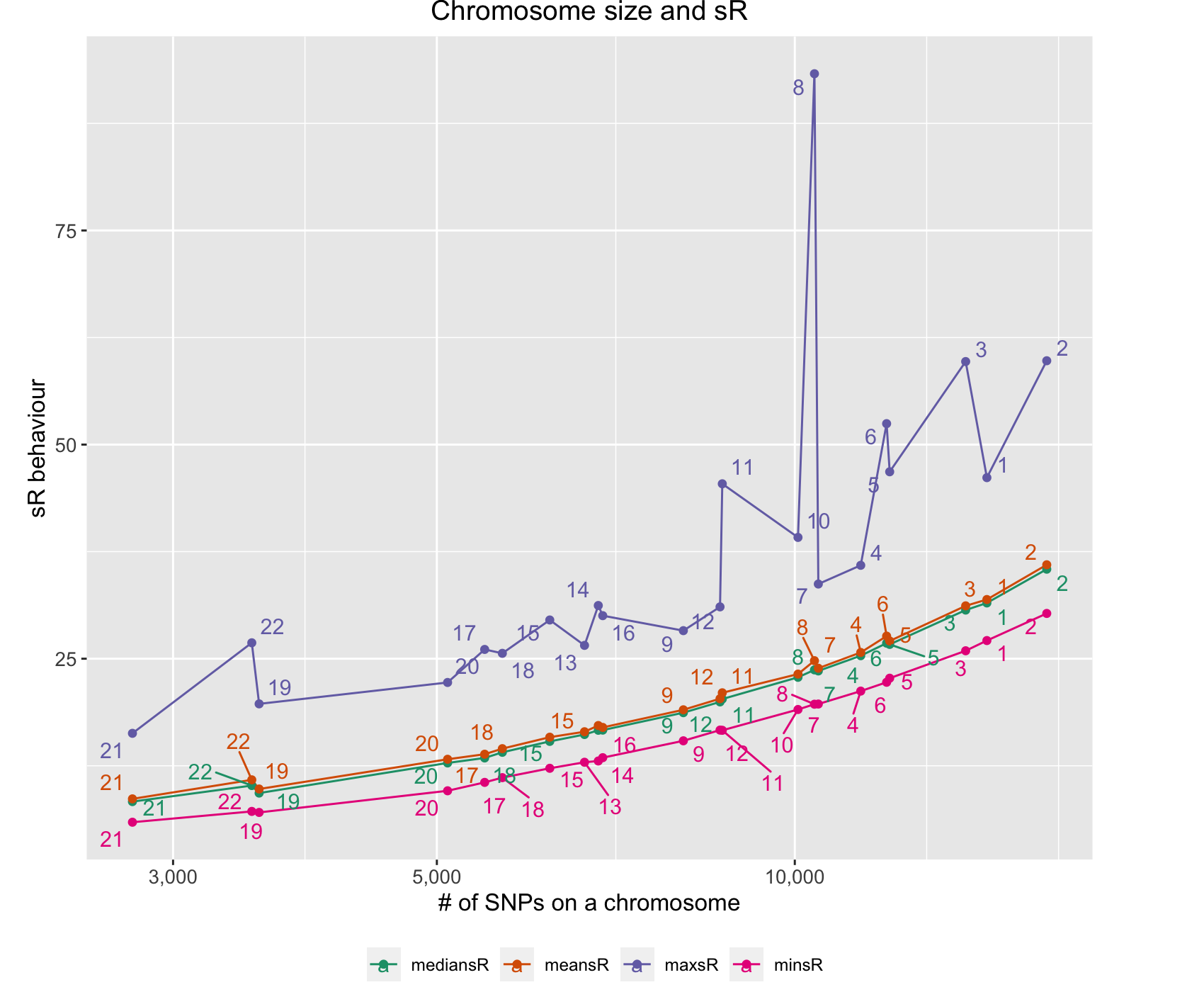}
\caption{A summary of multi-collinearity as a function of chromosome size for populations in Europe.}
\label{fig:EU_CHR}
\end{figure}

\begin{figure}[H]
\includegraphics[width=0.93\linewidth]{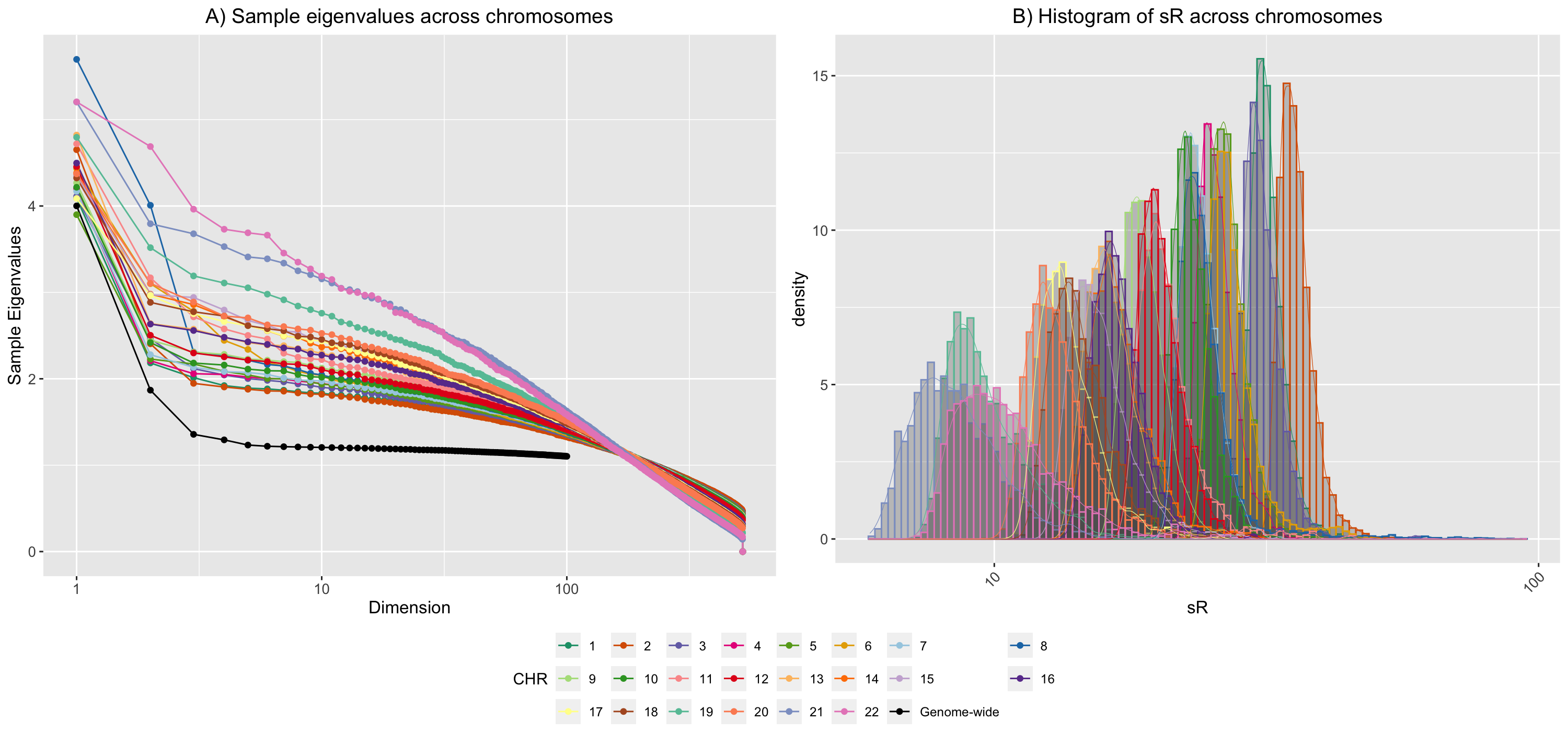}
\caption{Patterns of multi-collinearity measured by $\{sR_j\}_j$ in Europe.}
\label{fig:EU_summary}
\end{figure}

\begin{figure}[H]
\includegraphics[width=0.93\linewidth]{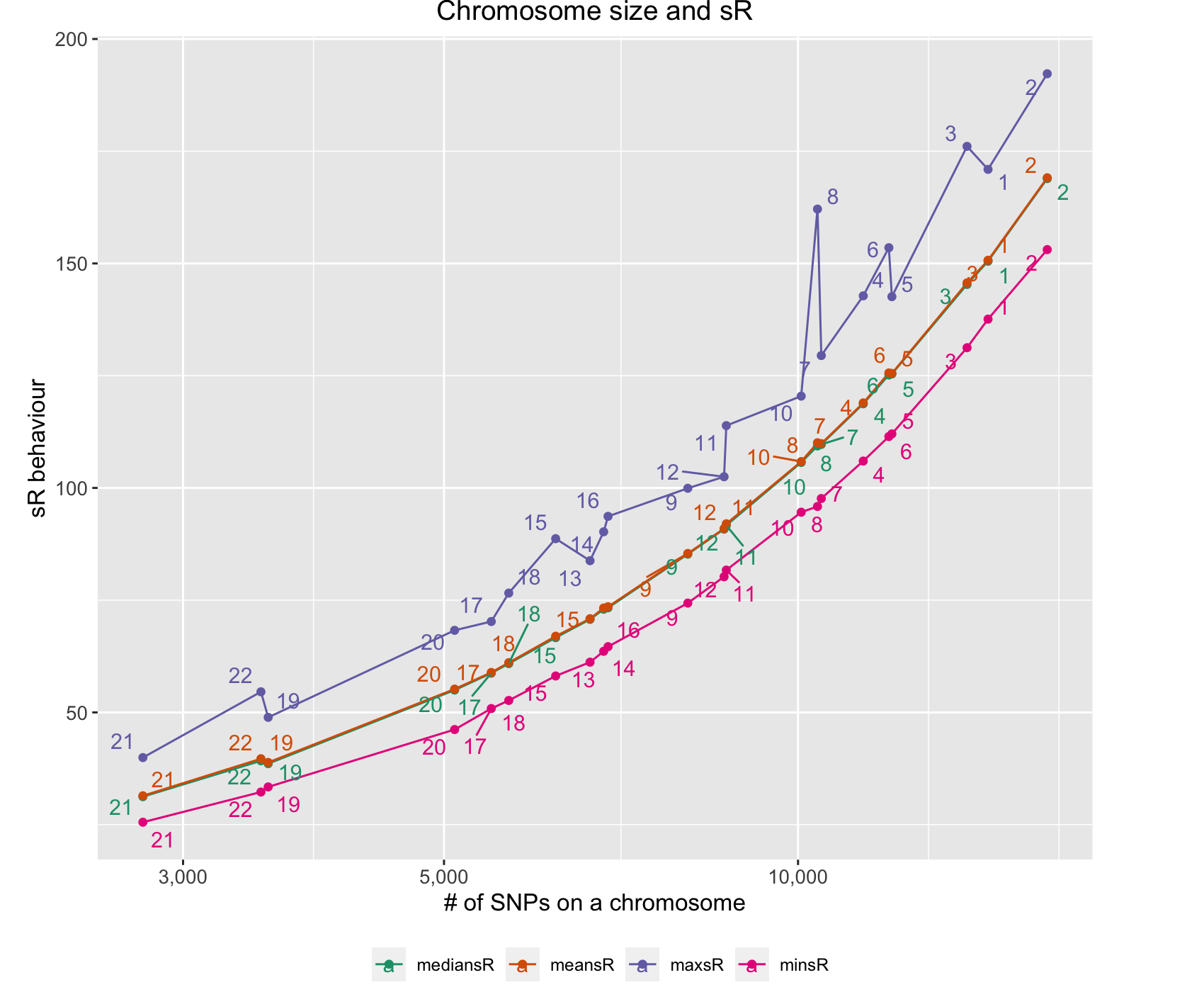}
\caption{A summary of multi-collinearity as a function of chromosome size for populations in South Asia.}
\label{fig:SA_CHR}
\end{figure}

\begin{figure}[H]
\includegraphics[width=0.93\linewidth]{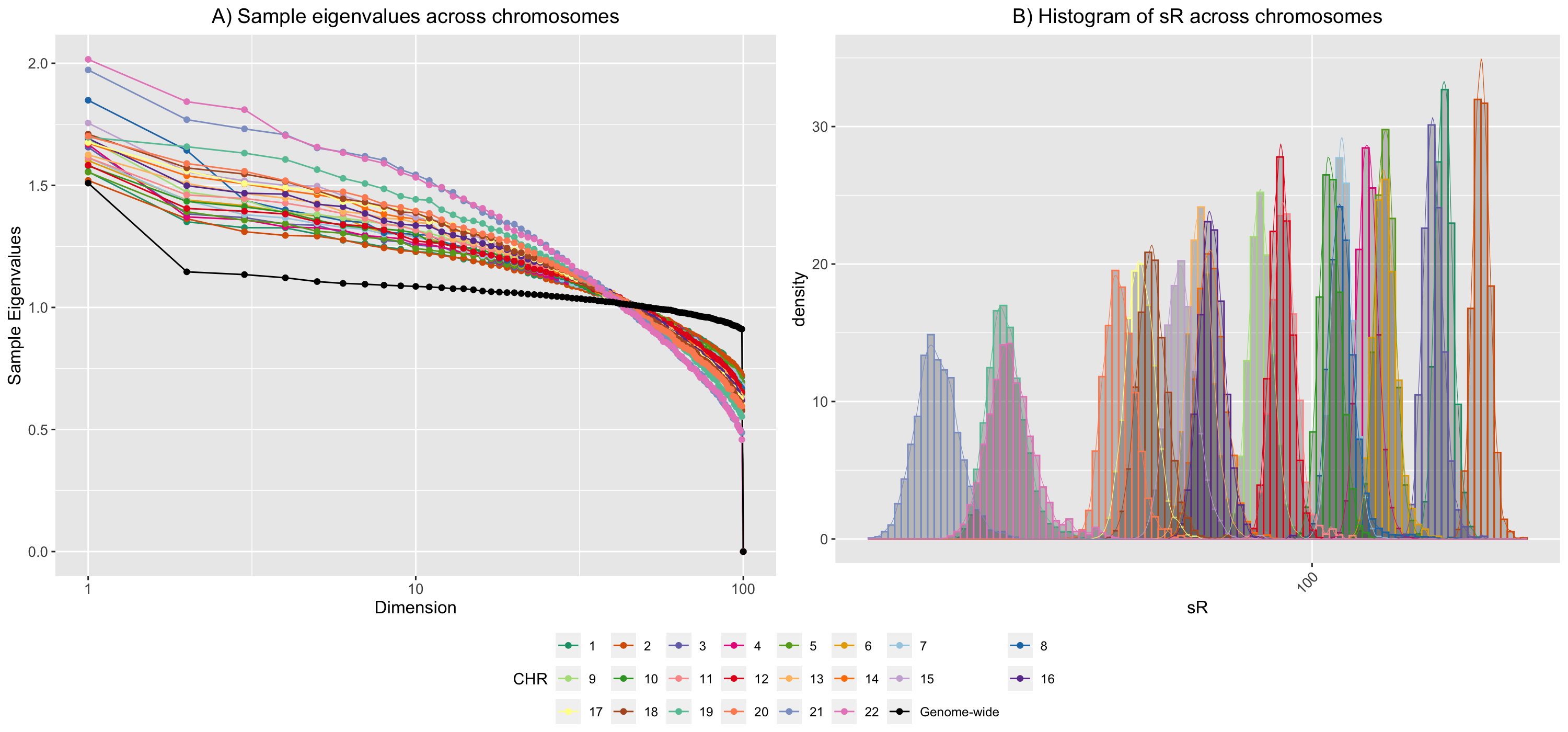}
\caption{Patterns of multi-collinearity measured by $\{sR_j\}_j$ in South Asia.}
\label{fig:SA_summary}
\end{figure}

\begin{figure}[H]
\includegraphics[width=0.93\linewidth]{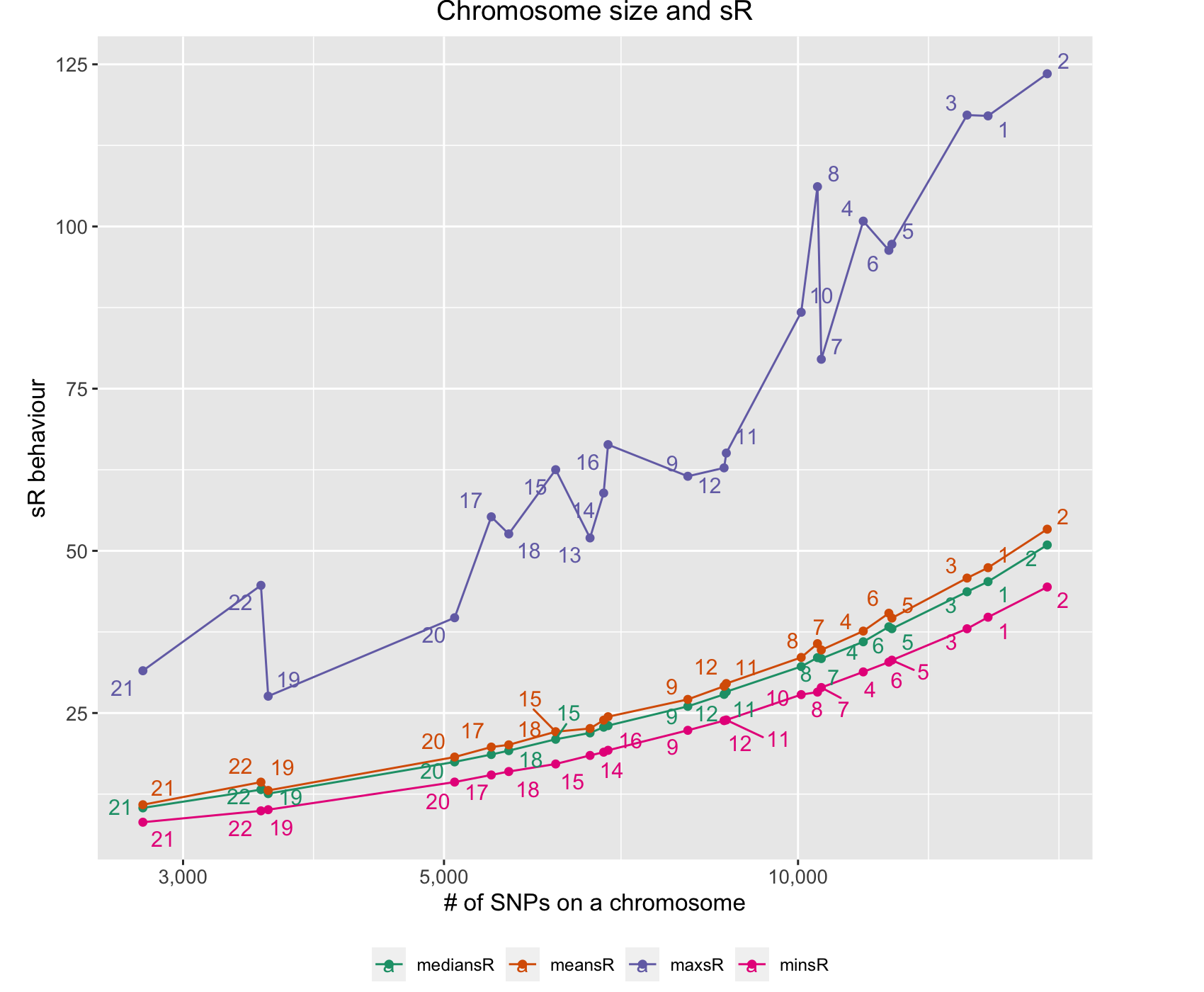}
\caption{A summary of multi-collinearity as a function of chromosome size for populations in Africa.}
\label{fig:AFR_CHR}
\end{figure}

\begin{figure}
\includegraphics[width=0.93\linewidth]{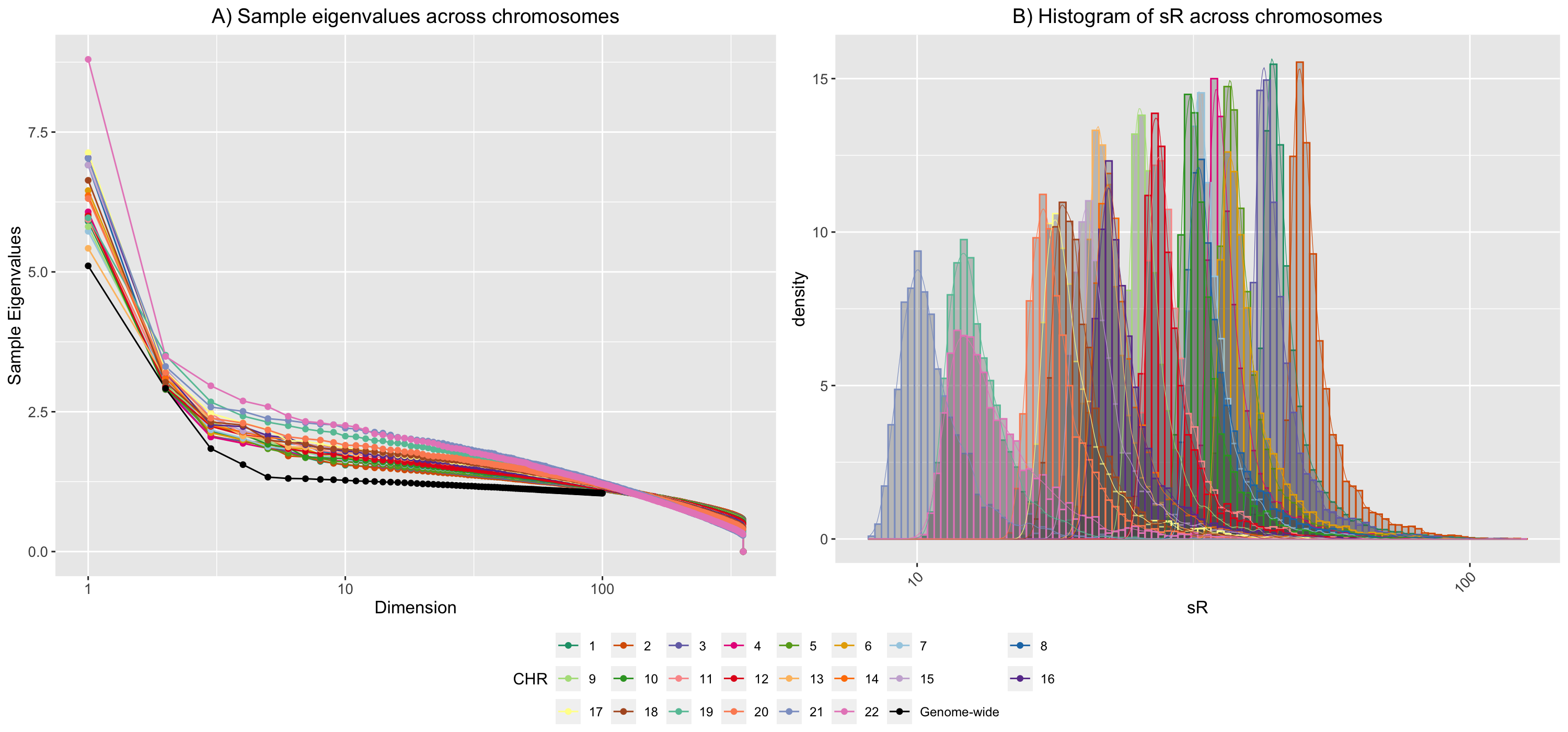}
\caption{Patterns of multi-collinearity measured by $\{sR_j\}_j$ in Africa.}
\label{fig:AFR_summary}
\end{figure}

\begin{figure}[H]
\includegraphics[width=0.93\linewidth]{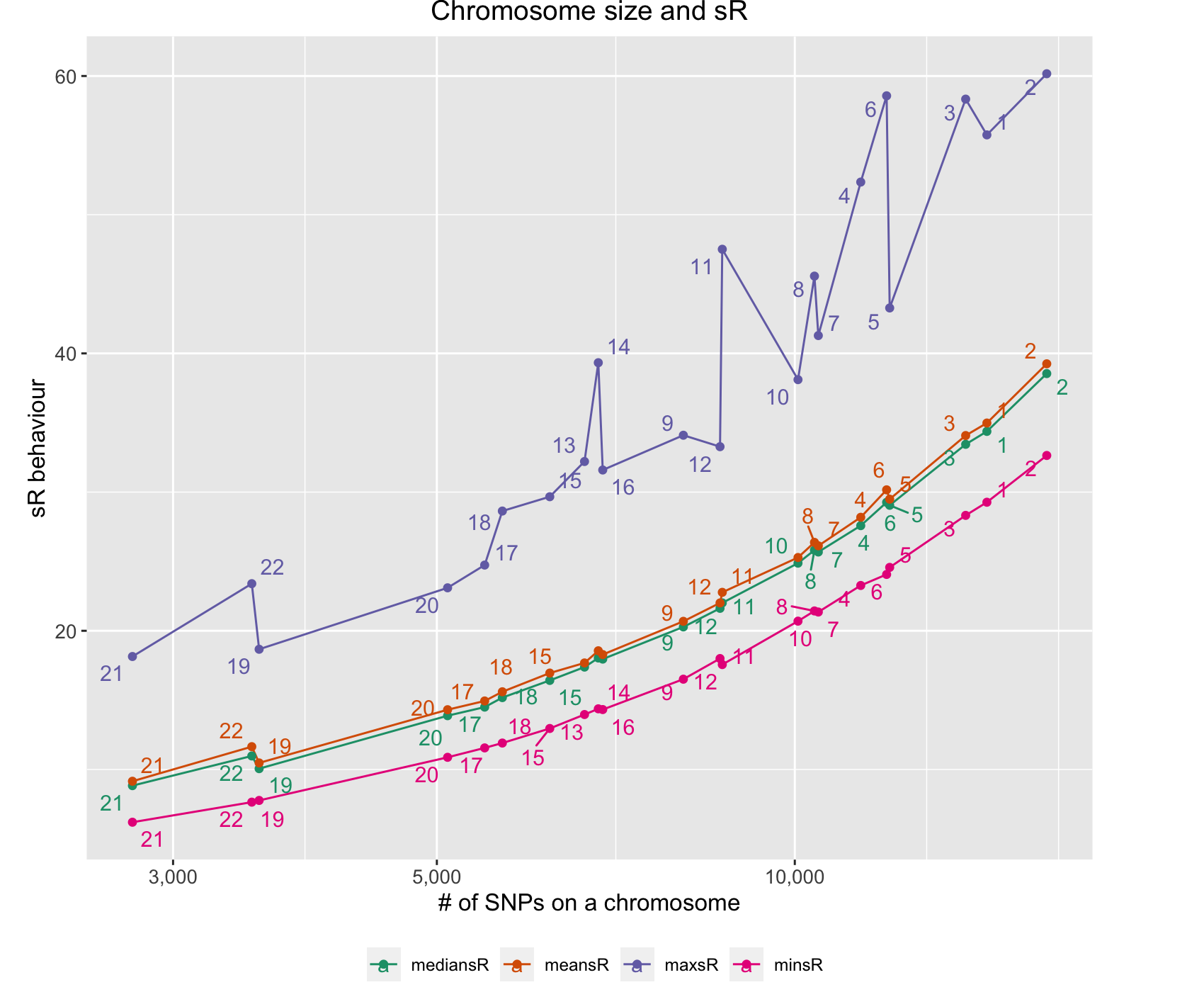}
\caption{A summary of multi-collinearity as a function of chromosome size for populations in East Asia.}
\label{fig:EA_CHR}
\end{figure}

\begin{figure}[H]
\includegraphics[width=0.93\linewidth]{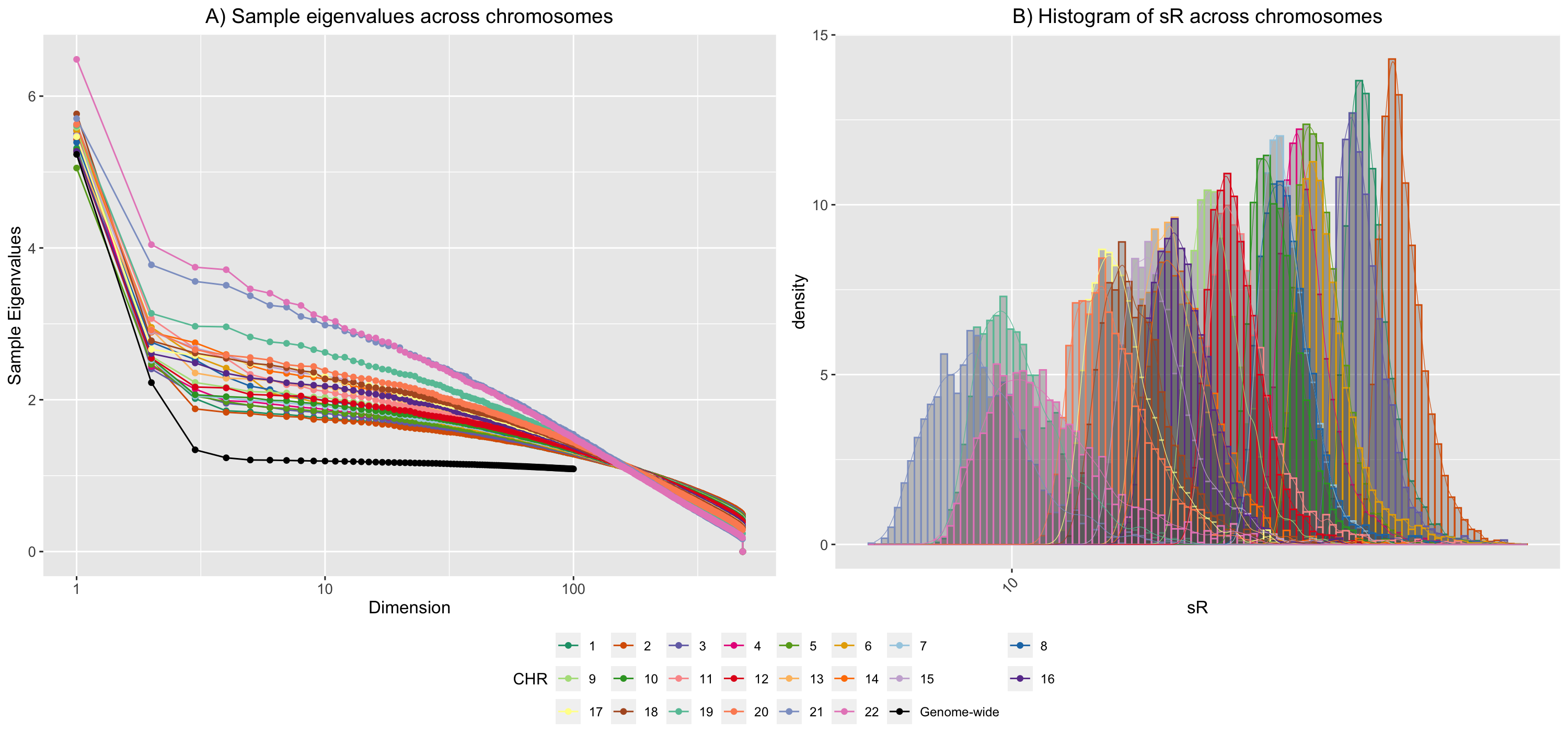}
\caption{Patterns of multi-collinearity measured by $\{sR_j\}_j$ in East Asia.}
\label{fig:EA_summary}
\end{figure}

\bibliographystyle{unsrtnat}
\bibliography{bibliography}  %%% Uncomment this line and comment out the ``thebibliography'' section below to use the external .bib file (using bibtex) .





% \end{thebibliography}

\end{document}